\documentclass[aps,amsfonts,nofootinbib]{revtex4-1}
\usepackage{epsfig}
\usepackage{graphicx}
\usepackage{amsmath}
\usepackage{color}
\usepackage{epsfig}
\usepackage{bbm}
\usepackage{pdfsync}
\usepackage{amsbsy}
\usepackage{mathrsfs}
\usepackage{caption}
\usepackage{subfigure}
\usepackage{float}
\newcommand{\bb}{\begin{equation}}
\newcommand{\ee}{\end{equation}}
\newcommand{\be}{\begin{equation}}
\newcommand{\eqb}{\begin{eqnarray}}
\newcommand{\eqf}{\end{eqnarray}}
\newcommand{\ba}{{\bar{a}}}
\newcommand{\bbb}{{\bar{b}}}
\newcommand{\bpa}{{\bar{\pi}}_a}
\newcommand{\bpb}{{\bar{\pi}}_b}
\newcommand{\bt}{{\bar{t}}}

\begin{document}

\title{Inflation without Inflaton: A Model for Dark Energy}
\author{H. Falomir }
\email{falomir@fisica.unlp.edu.ar}
\affiliation{Departamento de F\'{\i}sica,  Universidad Nacional de La
  Plata, La Plata, Argentina}
\author{J. Gamboa }
\email{jorge.gamboa@usach.cl}
\affiliation{Departamento de  F\'{\i}sica, Universidad de  Santiago de
  Chile, Casilla 307, Santiago, Chile}
 \author{P. Gondolo }
\email{paolo.gondolo@utah.edu}
\affiliation{Departament  of Physics,  University of  Utah, Salt  Lake
  City, Utah, USA }
\author{F. M\'endez }
\email{fernando.mendez@usach.cl}
\affiliation{Departamento de  F\'{\i}sica, Universidad de  Santiago de
  Chile, Casilla 307, Santiago, Chile}

\begin{abstract}

  The interaction between two  initially causally disconnected regions
  of  the  universe  is  studied using  analogies  of  non-commutative
  quantum  mechanics  and  deformation of  Poisson  manifolds.   These
  causally  disconnect   regions  are  governed  by   two  independent
  Friedmann-Lema\^{\i}tre-Robertson-Walker  (FLRW) metrics  with scale
  factors  $a$  and $b$  and  cosmological  constants $\Lambda_a$  and
  $\Lambda_b$, respectively. The causality is turned  on by positing a non-trivial
  Poisson bracket $[  {\cal   P}_{\alpha},  {\cal  P}_{\beta}  ] =
  \epsilon_{\alpha \beta}\frac{\kappa}{G}$,  where $G$  is  Newton's gravitational constant  and
  $\kappa $ is a dimensionless parameter.  The posited deformed Poisson bracket has an interpretation in terms of 3-cocycles, anomalies and Poissonian manifolds.
  The modified FLRW equations acquire an  energy-momentum tensor  from which we explicitly obtain
  the  equation of state parameter.   The   modified  FLRW  equations  are  solved
  numerically and the solutions  are inflationary or oscillating
  depending on the values of  $\kappa$. In  this model the  accelerating and decelerating  regime may be
  periodic.  The analysis  of  the  equation of state clearly shows  the
  presence of dark energy.  By completeness, the perturbative solution
  for $\kappa \ll1 $ is also studied.
  \end{abstract}

%\pacs{PACS numbers:12.60.-i, 11.30.Cp, }
\date{\today}

\maketitle

%%%%%%%%%%%%%%%%%%%%%%%%%%%%%%%
%%%%%%%%%%%%%%%%%%%%%%%%%%%%%%%
\section{motivating the problem}
%%%%%%%%%%%%%%%%%%%%%%%%%%%%%%%
%%%%%%%%%%%%%%%%%%%%%%%%%%%%%%%

Understanding the origin and behavior  of dark matter and dark energy
poses one of the most important challenges of today's physics, and its
solution could require new radical ideas.

Standard cosmology rests on the cosmological principle, the assumption
that   the   universe   is   homogeneous  and   isotropic   on   large
scales. However in the Big  Bang era, approximately 13.8 billion years
ago, when the Universe violently expanded from a very high density and
temperature state, the cosmological principle conditions  were not fulfilled  because of the
extraordinarily   non-homogeneous  and   anisotropic  nature   of  this
expansion. The released energy, then,  was redistributed in such a way
that causally disconnected sectors were formed \cite{Linde}.

After this  extremely short period of  time our known laws  of physics
apply  and one  can speculate,  for  example, about  the formation  of
topological   defects  which   break  the   large  scale   homogeneity
\cite{Kibble,Vilenkin,monopole},  as domain  walls, cosmic  strings or
monopoles, of which no visible sign has been found.

This  led to  the assumption  of a  period of  \emph{cosmic inflation}
\cite{Starobinsky,guth,linde,steinhardt,linde1,linde2,linde3,mukhanov,kazana,sato}
during   which  the   universe  grew   exponentially,  smoothing   out
inhomogeneities   inside the  cosmological  horizon,  the
boundary of  our observable causal  patch of the  universe.  Inflation
ends  in a  reheating phase  where  the standard  model particles  are
produced and,  as temperature decreases, quantum  fluctuations explain
the formation of galaxies and the current large-scale structure of the observable
Universe.

After the cosmic inflation, most of  the evolution of the Universe has
been dominated by  matter and radiation. But evidence  coming from the
red-shift  of   Type  Ia   supernovae  \cite{Type-Ia}  and   from  the
fluctuations in  the cosmic  microwave background  \cite{WMAP} suggests
that   our  Universe   is  presently   in  a   phase  of   accelerated
expansion.  Lambda Cold  Dark  Matter ($\Lambda$CDM)  is the  standard
cosmological model describing this situation, with approximately 4.9\%
of ordinary (baryonic) matter, 26.8\% of (cold) \emph{dark matter} and
68.3\%   of    \emph{dark   energy}   ($\Lambda$   stands    for   the
\emph{cosmological  constant}), compatible  with a  flat space  with a
critical                         total                         density
$\rho_{tot}=\rho_{matt}+{\Lambda}/8\pi G\approx  3{H}^2/8\pi G$ (where
$H$ the Hubble parameter). The origin of dark matter and dark energy is
still unknown.

The universe in this model is  described by the Friedmann--Lema\^itre--Robertson--Walker
(FLRW)  metric
\begin{equation}\label{RW}
    {ds}^2=- {dt}^2+{a(t)}^2 {R_0}^2  \left\{ \frac{{dr}^2}{1-k r^2} +
      r^2 {d\Omega}^2\right\}\,,
\end{equation}
where $a(t)$ is the (dimensionless) time-dependent \emph{scale factor}
of spatial sections,  $R_0$ is the present length scale (if  $a(0)=1$ at  the
present time $t=0$), $r$ is  the (dimensionless) radial coordinate and
$k$  is the  curvature of  spatial sections,  being $k=0$  for a  flat
space.  For this  geometry,  Einstein's equations  reduce  to the  two
Friedmann's equations \cite{frieman}
\begin{equation}\label{Friedmann}
    \begin{array}{c}\displaystyle
      H(t)^2:=  \left(\frac{\dot{a}({t})}{a(t)}\right)^2 =  \frac{8\pi
      G}{3}\, \rho+\frac{\Lambda}{3}-\frac{k}{{R_0}^2{a(t)}^2}\,,
      \\ \\ \displaystyle
      \frac{\ddot{a}(t)}{a(t)}   =  -\frac{4\pi   G}{3}\left(  \rho+3p
      \right)+\frac{\Lambda}{3}\,,
    \end{array}
\end{equation}
where \emph{dots} refer to time-derivatives.
The role  of $\Lambda$ is  clear from the previous  equations. Indeed,
for a $\Lambda$-dominated era -- the dark energy dominated epoch -- we
have  that asymptotically  $a(t) \sim  e^{t {\sqrt{\Lambda/3}}}$.  The
physical origin of the cosmological constant  is troublesome,
on  the other hand.   For example, the identification of $\Lambda$ with the vacuum energy of
the various species of particles of the standard model, evaluated with
a cut-off  of the order  of the Planck scale,  leads to a  mismatch of
around 120  orders of  magnitude when  compared with  observable data,
while  an  exact  supersymmetric  field theory  predicts  a  vanishing
result.  If SUSY  is broken -- a  natural way out to the  problem -- a
\emph{fine tuning} is necessary  to approximate the experimental value
of $\Lambda$.

This  is  the  \emph{cosmological  constant  problem}  \cite{Weinberg,
  Zeldovich,  Carroll,  peebles}, one  of  the  most significant  open
problems in fundamental physics. Models to solve this puzzle have been
formulated in which  $\Lambda$ is related with  the vacuum expectation
value  of  the  energy  density   of  dynamical  light  scalar  fields
\cite{wetterich}  with local  minima  in the  potential energy,  which
would produce  \emph{phase transitions} as the  temperature decreases,
or with a sufficiently small slope to  produce a slow roll down to the
minimum of the potential.

Different scenarios  explore the possibility that  cosmic acceleration
could be  described by  higher-dimensional theories  \cite{dvali}.  It
has  also  been  argued  \cite{Chang:2001bm}  that  Quantum  Mechanics
combined with  Einstein's theory  would require  a kind  of nontrivial
uncertainty relations at the Planck scale, which impose effective short
distance or large momentum cut-offs,  since the attempt to localize an
event with extreme  precision would demand an energy  that would lead
to a gravitational collapse.

\smallskip

In the present work we consider  a model where patches of the universe
-- causally disconnected after the inflation era --  have  independently
evolved  to a  dark  energy dominated  era  according to  Einstein's
equations. A  tiny effective interaction  is considered, which  can be
interpreted  as  a  relic  of  a  primordial  non-trivial  uncertainty
relation.

Technically, a mechanism that might explain how our universe evolved
to the  conditions we know today  --   an accelerated expansion state -- is based on a deformed Poisson bracket structure for
the    metrics   describing    the    patches    of  initially
causally-disconnected components.

Indeed, deformed Poisson bracket  algebras, which describe phase-space
noncommutative geometries \cite{connes,  flato, catta, wilde, fedosov,
  flato1, kosen},  usually imply nontrivial  interactions \cite{ncqm}.
As       an       example,        consider       the       Hamiltonian
$\mathcal{H}_0=\frac{1}{2}\,  \pi_i \pi_i$  (sum  is  implied) in  two
dimensions,         with        phase         space        coordinates
$\{x^i,\pi_j\}_{\{i,j\}\in\{1,2\}}$   satisfying   the   non-canonical
Poisson brackets
\begin{equation}
\label{PB2}
    \left\{x^i, x^j\right\}  = 0\,, \quad \left\{x^i,  \pi_j\right\} =
    \delta^i_j\,, \quad \left\{\pi_i, \pi_j\right\} = B \epsilon_{i
      j}\,. \quad
\end{equation}
Here  $B$   is  a  constant   and  $\epsilon_{ij}$  is   the  totally
antisymmetric tensor.

This system describes the Landau model as can be seen by performing
the change of variables in the momentum
sector $\pi_j = p_j + \frac{B}{2} \, \epsilon_{j k} x^k $ (the so called {\it Bopp's shift}).  The system
is now described by the Hamiltonian
 \begin{equation}
\label{PB3}
    \mathcal{H} = \frac{1}{2}\, \left( p_i + \frac{B}{2}\, \epsilon_{i
        k} x^k \right)^2\,,
\end{equation}
while the Poisson bracket structure is the canonical one, namely
\begin{equation}
\label{PB1}
  \left\{x^i,  x^j\right\}  =  0\,, \quad  \left\{x^i,  p_j\right\}  =
  \delta^i_j\,, \quad \left\{p_i, p_j\right\} = 0\,.
\end{equation}

Let  us  just  mention  that  performing a  linear  Bopp's  shift  is
equivalent  to using  the $\star$-Moyal  product when  the deformation
parameters    are    constant.     Otherwise,   the    more    general
$\star$-Kontsevich product or an  $\hbar$-expansion should be employed
\cite{kosen}.

\smallskip

This paper  is organized as follows.  In section II, we  implement the
main idea  by extending the  FLRW metric to  a universe with  two FLRW
metrics -- that is with two scale factors $a(t)$ and $b(t)$ -- coupled
by rules  resembling the  case of  the Landau  problem, which  will be
explained  below.  Basically,  the idea  is to  consider the  standard
cosmological  model  as  a  Hamiltonian  system  formally  similar  to
classical mechanics with the second metric  as a new degree of freedom
coupled through a  Landau-like mechanism.  In section  III the coupled
FLRW  equations  are  solved,  first by  using  numerical  methods  in
subsection  (a) and  then through a first-order perturbative expansion in
subsection (b).
These solutions  show inflation at early times  but
another behavior emerges at  late times. The inflation at early times
is  an  effect  that   becomes  manifest  assuming  that  cosmological
constants   in  the   two   different  patches   satisfy  a   relation
$\Lambda_a \ll \Lambda_b$.  In section IV, the interpretation in terms
of dark  energy is given  for the  solutions previously found, and the
last section is devoted to discussion and conclusions.

%%%%%%%%%%%%%%%%%%%%%%%%%%%%%%
%%%%%%%%%%%%%%%%%%%%%%%%%%%%%%
\section{The model}
%%%%%%%%%%%%%%%%%%%%%%%%%%%%%%
%%%%%%%%%%%%%%%%%%%%%%%%%%%%%%
Following the  analogy with  the Poisson manifold  deformation (Landau
problem)  described above,  we  consider a  universe  with two  scale
factors $a(t)$ and $b(t)$ and we $\underline{\it posit}$ the following
deformed Poisson bracket for the conjugate momenta $\pi_a$ and $\pi_b$ of $a$
and $b$:
 \bb \left\{ \pi_\alpha , \pi_\beta \right\} =
\epsilon_{\alpha \beta} \, \theta.
 \label{d2}
 \ee
 Here, indices  $\alpha,\beta \in \{1,2\}$ label  scale factors --
 that is, $a_1\equiv a,~ a_2\equiv b,~\pi_1\equiv \pi_a,~ \pi_2 \equiv
 \pi_b $  -- while  $\epsilon_{\alpha \beta}$  is the  two dimensional
 Levi-Civita  tensor.

 The $\theta$ parameter can be  chosen with dimensions of (energy)$^2$
 if  $a_\alpha$ has  dimensions  of  (energy)$^{-1}$ and  $\pi_\alpha$
 dimensions of energy (see the Appendix). However, we must note  that,
 contrarily to the  Landau problem, $\theta$ should  not be identified
 with an external  magnetic field and its value should  be fixed using
 different physical arguments.

The remaining Poisson brackets are the standard ones, {\it i.e.}
\begin{equation}
\label{usual}
\left\{ a_\alpha, a_\beta \right\} = 0\,, \quad
\left\{ a_\alpha,\pi_\beta \right\} = \delta_{\alpha \beta}.
\end{equation}

Before continuing  with the technical  discussion, let us  introduce a
useful parameterization  for $\theta$. Indeed,  since it is  a constant
for the  two metrics  under consideration in  this universe,  it seems
natural to  define $  \theta =\frac{\kappa}{G}$,  where $\kappa$  is a
dimensionless parameter and $G$ is Newton's constant.
From this point of view,
the  $G \to \infty$ limit  with fixed  $\kappa$ (formally
equivalent to $\kappa \to 0$ with fixed $G$) 
would correspond to a universe with two causally disconnected
patches.

In fact, the  $G\to \infty$ limit closely resembles the
tensionless limit in string theory  \cite{marti}. In string theory the
tension of  the string $T$ and  the Regge slope $\alpha'$  are related
through   $T  \propto   \frac{1}{\alpha'}$,   the  tensionless   limit
corresponds  to  $\alpha'  \to   \infty$,  generating  the  so  called
ultra-local limit,  where every point  in the string  evolves in a causally
disconnected manner from  the  rest  of  the points  on  the  same  string
(parenthetically,  many efforts  were  devoted to  the  study of  such
scenario   as  can   be   seen,  for   example,  in   \cite{isham}. See also  \cite{marc}).  However  we stress  the difference
with our approach where no relic  of spatial dependence appears in the
metric, as a consequence of the cosmological principle.

For  the total  Hamiltonian of  the  model presented  here, under  the
conditions set out above, we take\footnote{We are taking the same time coordinate for both sectors. We recall that Friedmann's equations can be derived from the effective Lagrangian
\begin{equation}\nonumber
    L[a,\dot{a},N]:=\frac{1}{G}  \left[ \frac{a \, {\dot{a}}^2}{2 N}+ \frac{\Lambda}{6}\, N  a^3  -\frac{k}{2}\, N a\right],
\end{equation}
where $N(t)$ is an auxiliary variable which ensures the time-reparameterization invariance of the action. Indeed, the second line in  \eqref{Friedmann} is the equation of motion for $a(t)$ if we choose $N(t)\equiv 1$, while the equation for $N(t)$ imposes the first line in \eqref{Friedmann} as a constraint on the system. The corresponding Hamiltonian, with $p_a= {a \dot{a}}/{G N}$ and $p_N=0$, reads 
\begin{align}
    & H(p_a,a,p_N,N) =N \mathcal{H}(p_a,a),
    \nonumber
\intertext{where}
    & \mathcal{H}(p_a,a) = \frac{1}{2} \left[ {G}\, \frac{{p_a}^2}{a}-\frac{\Lambda}{3 G}\, a^3 + \frac{k}{G}\, a \right],
    \nonumber
\end{align}
and the constraint implies $\mathcal{H}= 0$ on the physically acceptable trajectories.}
\bb
H = N\left[  \frac{G\pi_a^2}{2 a} + \frac{{G \pi}_b^2}{2b} +
\frac1{2G}\left(k_a a -   \frac{\Lambda_a}{3}\,a^3 + k_b b -
\frac{\Lambda_b}{3}\,b^3\right)\right],
\label{dP}
\ee where  $k_a$ and $k_b$ are  the spatial curvatures of  the patches
described by  the scales $a(t)$ and  $b(t)$, respectively.

The equations of motion derived from  this Hamiltonian with the Poisson
bracket structure  defined in (\ref{d2}) and  (\ref{usual}) turn out
to be
\begin{eqnarray}\label{eqs-ham}
{\dot a} &=&G\frac{\pi_a}{a},
\label{eoma}
\\
{\dot b} &=&G\frac{\pi_b}{b},
\label{eomb}
\\
\dot{\pi}_a &=& G\frac{\pi_a^2}{2a^2} +\frac{\Lambda_a a^2-k_a}{2G} + \kappa
\frac{ \pi_b}{b},
\label{eompa}
\\
\dot{\pi}_b &=& G\frac{\pi_b^2}{2b^2} +\frac{\Lambda_b b^2 -k_b}{2G} -
\kappa \frac{\pi_a}{a},
\label{eompb}
\end{eqnarray}
where we have used the reparameterization $Ndt\to dt$ or, equivalently,
we have taken $N=1$ at the end of the derivation.

The constraint {$\dot{p}_N = 0$} derived from this Hamiltonian (a consequence of time-reparameterization invariance of the effective action) turns out to be
\bb
\frac{\pi_a^2}{a} + \frac{{ \pi}_b^2}{b} +
\frac1{G^2}\left(k_a a -   \frac{\Lambda_a}{3}\,a^3 + k_b b -
\frac{\Lambda_b}{3}\,b^3\right)= 0.
\label{vin}
\ee
Notice that this constraint is independent of $\kappa$ and so it applies to our model even in the canonical Poisson brackets case.

The equations of motion (\ref{eoma}) and (\ref{eomb}) can be used to write
the momenta  equations (\ref{eompa})  and  (\ref{eompb})  as second  order differential
equations, and also to bring equation (\ref{vin}) to the \emph{standard} form.
In so doing we find the following set of equations
\eqb
 2\,  \frac{{\ddot  a}}{a}  +  \left(\frac{{\dot  a}}{a}\right)^2  &=&
 \Lambda_a - \frac{k_a}{a^2} + 2\kappa\,
 \frac{{\dot b}}{a^2},
 \label{moei11}
\\
 2\, \frac{{\ddot b}}{b}+ \left(\frac{{\dot b}}{b}\right)^2 &=&
\Lambda_b -\frac{k_b}{b^2} - 2\kappa \frac{ {\dot a}}{b^2},
  \label{moei22}
\\
\label{connew}
a {\dot  a}^2+b{\dot b}^2  &=&
\frac{\Lambda_a}{3} a^3  -k_a a + \frac{\Lambda_b }{3}  b^3 -k_b b.
\eqf
They contain all the dynamical information about the model and  show that
the evolution  of the  scale factor  of one patch  is modified  by the behavior of the
scale factor of  the other patch. We can venture  an interpretation here
in terms of gravitational bubbles. Indeed, each scale factor describes
one of the bubbles and they evolve under a sort of interaction induced
by (\ref{d2}).

The previous system of differential  equations  can also  be  derived  from a  canonical  Poisson
structure by performing a Bopp's  shift in the momenta $\pi_\alpha$,
in complete analogy with  non-commutative quantum mechanics.  That is,
we can perform a change of  variables in the form
 \bb \pi_\alpha  =
   p_\alpha + \frac{\theta}{2}\epsilon_{\alpha \beta} a_\beta
  =
   p_\alpha  +  \frac{\kappa}{2G}   \epsilon_{\alpha  \beta}  a_\beta,
   \nonumber
 \ee
where $p_\alpha$  are the  canonical momenta  ($\{a_\alpha,p_\beta\} =
\delta_{\alpha\beta}$).

The Hamiltonian in these variables turns out to be
\begin{equation}
\label{hamusual}
H=N \left[
\frac{G}{2a} \left( p_a + \frac{\kappa}{2G}b   \right)^2 +
\frac{G}{2b} \left( p_b - \frac{\kappa}{2G}a   \right)^2 +
\frac{1}{G}\left(k_a a -   \frac{\Lambda_a}{3}\,a^3 + k_b b -
\frac{\Lambda_b}{3}\,b^3\right)
\right].
\end{equation}
The corresponding Hamilton's equations  of motion are
\begin{eqnarray}
{\dot a} &=&\frac{G}{a} \left( p_a+\frac{\kappa}{2G} b \right),
\label{eoma2}
\\
{\dot b} &=&\frac{G}{b} \left( p_b- \frac{\kappa}{2G} a \right),
\label{eomb2}
\\
\dot{p}_a &=& \frac{G}{2a^2} \left( p_a+\frac{\kappa}{2G} b \right)^2+
              \frac{\Lambda_a a^2-k_a}{2G} +
\frac{\kappa}{2b} \left( p_b - \frac{\kappa}{2G} a \right),
\label{eompa2}
\\
\dot{p}_b  &=&   \frac{G}{2b^2}  \left(  p_b  -   \frac{\kappa}{2G}  a
              \right)^2+ \frac{\Lambda_a a^2-k_a}{2G} -
\frac{\kappa}{2a} \left( p_a + \frac{\kappa}{2G} b \right),
\label{eompb2}
\end{eqnarray}
where the  gauge $N=1$  has been chosen again. Of course, the constrained system of second  order differential
equations derived from here is also given by  (\ref{moei11}), (\ref{moei22}) and (\ref{connew}).

Finally note  that it  is possible  to identify  the right-hand sides  of equations
(\ref{moei11}) and (\ref{moei22}) with  an energy-momentum tensor that
is covariantly conserved.   Indeed, let us consider  the $a$-sector of
the model,  that is the patch  of the universe described  by the scale
factor  $a$. The FLRW Einstein tensor $G_{\mu\nu}$ for the $a$-patch
has time
component $G^{0}_{\,\,\,0} =-3({\dot a}^2 +k_a)/a^2$ and space components 
$G^{1}_{\,\,\,1}=G^{2}_{\,\,\,2}=G^{3}_{\,\,\,3}=-(2a\ddot{a}+\dot{a}^2+k_a)/a^2$. Then
equations (\ref{moei11}) and (\ref{connew}) can be written as the Einstein equations for the $a$-patch
\eqb
G_{\mu\nu} = 8 \pi \,G\,T_{\mu\nu} ,
\eqf
provided
\eqb
T^0_{\,\,\,0} &=& \frac{\Lambda_a}{8\pi G} + \frac{b^3}{a^3} \bigg[ \Lambda_b - \frac{3 ( {\dot b}^2 + k_b)}{b^2} \bigg],
\nonumber
\\
T^1_{\,\,\,1} &=&T^2_{\,\,\,\,2}= T^3_{\,\,\,\,3}= -\frac{\Lambda_a}{8\pi G} - \frac{\theta}{4\pi}\,\frac{{\dot b}}{a^2}.
 \label{conser}
\eqf
Notice that the term in square brackets in $T^0_{\,\,\,0}$ is $G^{0}_{\!(b)\,0} + \Lambda_b \delta^{0}_{\,\,\,0}$, with $G^{0}_{\!(b)\,0}=-3({\dot b}^2 +k_b)/b^2$, and thus vanishes when for $\theta=0$ the $a$ and $b$ patches evolve independently. Energy-momentum covariant conservation then implies that
\bb
T^\mu_{\,\,\,\,\,\nu;  \mu}= 0  \,\,\Rightarrow \frac{3{\dot  b}}{a^3}
\left( 2  b {\ddot b}  + {\dot b}^2 +k_b  -\Lambda_b b^2 +  2 \kappa\,
  {\dot a} \right)=0.
\ee
This  condition is  just the equation  of motion for $b(t)$, Eq.\ \eqref{moei22}.   This is  a
self-consistency property, intrinsic to the model proposed here. Notice that, in this way, each sector appears as a kind of  \emph{local} source of the other. Moreover, the effective density, which does not depend on $\kappa$, is induced by the time-reparameterization invariance through the constraint in Eq.~\eqref{vin}, while the effective pressure, proportional to $\kappa$, is a consequence of the assumed noncommutativity, Eq.~\eqref{d2}. Notice also that the sign of the effective pressure depends on the behavior of the scale $b(t)$ (expansion or contraction) of the second sector.

The realization of our model in a more fundamental theory is not unique and
deserves a careful analysis. A possible  picture of the model could be
the effective description  of two regions of  the universe, originally
disconnected (as, for  example, regions separated by domain  walls as
mentioned in  the introduction).   It might also be possible  to have an
interpretation in terms of two universes,\footnote{This point of view has
  been defended  vigorously by Linde, see  \cite{Linde} and references
  therein.}   albeit   initially   disconnected,   such   that   this
disconnection is  broken at a certain  time through the  modification of
the canonical brackets. Another interpretation is simply in terms of a
universe with two metrics, {\bf namely a bimetric theory, which does not suffer from the 
usual instabilities and in this sense our model resembles the results presented  by Freidel et al \cite{freidel1,freidel2}}.

Finally, and  this is the  idea we  explore now, the  possibility that the effect of
the \emph{dark energy} responsible for the accelerated expansion  of our universe
at the present time could  be encoded in a second metric interacting
in the way we have explored here is very attractive.

%%%%%%%%%%%%%%%%%%%%%%%%%%%%%%%%%%%
%%%%%%%%%%%%%%%%%%%%%%%%%%%%%%%%%%%
\section{Solutions of  the generalized FLRW equation}\label{NumSol}
%%%%%%%%%%%%%%%%%%%%%%%%%%%%%%%%%%%
%%%%%%%%%%%%%%%%%%%%%%%%%%%%%%%%%%%

In order to  study the properties of the model  proposed here, we will
analyze  the  solutions  of  the  equations  of  motion  in  different
regimes. The perturbative regime is defined  by $\kappa \ll 1$. We
will show  that this analysis can  be done consistently only  at early
times in the evolution of the universe.

For  late times,  instead,  numerical solutions  of  the equations  of
motion will  be useful.  For  this case,  instead of using  the second
order set of  equations (\ref{moei11}) and (\ref{moei22}),  it will be
convenient  to  solve  directly  the  Hamiltonian  system  defined  in
equations (\ref{eoma}) to (\ref{eompb}).

For  the analysis,  it  is convenient  to define  a  scale $\mu$  with
dimensions of  energy, that  is $[\mu]=+1$,  such that  the quantities
$\bar{t} \equiv \mu\,t,~\bar{a}\equiv \mu\,a,~\bar{b} \equiv \mu\,\,b$
are     dimensionless.      For     the     momenta      we     define
$\bar{\pi}_\alpha  \equiv   \mu\,G\,\pi_\alpha$  (or  in   terms  of
canonical variables $\bar{p}_\alpha \equiv \mu\,G\,p_\alpha$).
Finally, cosmological  constants can  be rescaled
also and we define $\lambda$ such that
$$
\sin   \lambda  \equiv   \frac{\Lambda_b}{\mu^2},~~~~~~~~\cos  \lambda
\equiv \frac{\Lambda_a}{\mu^2},
$$
so $ \mu = (\Lambda_a ^2 + \Lambda_b^2)^\frac{1}{4} $.

In terms  of these  dimensionless quantities,  the set  of first-order
dynamical equations reads
\eqb
\frac{d {\bar a}}{d {\bar t}}&=& \frac{{\bar \pi}_a}{{\bar a}},
\label{fo1}
\\
\frac{d {\bar b}}{d {\bar t}}&=& \frac{{\bar \pi}_b}{{\bar b}},
\label{fo2}
\\
\frac{d  {\bar  \pi}_a}{d  {\bar t}}&=&  \frac{{\bar  \pi}^2_a}{2{\bar
    a}^2}  + \frac{{\bar  a}^2  \cos \lambda  -k_a}{2} +  \frac{\kappa
  {\bar \pi}_b}{{\bar b}},
\label{fo3}
\\
\frac{d  {\bar  \pi}_b}{d  {\bar t}}&=&  \frac{{\bar  \pi}^2_b}{2{\bar
    b}^2} + \frac{{\bar b}^2 \sin \lambda -k_b}{2} -\frac{\kappa {\bar
    \pi}_a}{{\bar a}},
\label{fo4}
\eqf
while the second order dimensionless system is
\eqb
 2\, \bar{a}\, \frac{d^2 \bar{a}}{d\bar{t}^2} + \left(
   \frac{d\bar{a}}{d\bar{t}} \right)^2
 &=& \bar{a}^2\,\cos\lambda - k_a + 2\kappa \, \frac{d\bar{b}}{d\bar{t}},
 \label{sec1}
\\
2\,    \bar{b}\,\frac{d^2     \bar{    b}}{d\bar{t}^2}     +    \left(
  \frac{d\bar{b}}{d\bar{t}} \right)^2
 &=& \bar{b}^2\,\sin \lambda - k_b - 2\kappa\, \frac{d\bar{a}}{d\bar{t}}.
\label{sec2}
\eqf
Finally, the dimensionless form of the constraint is
\begin{equation}
\label{constdimless}
\bar{a}\, \left(  \frac{d\bar{a}}{d\bar{t}} \right)^2 + \bar{b}
\left( \frac{d\bar{b}}{d\bar{t}} \right)^2 = \bar{a}^3 \,
\frac{\cos\lambda}{3}   +   \bar{b}^3\,   \frac{\sin\lambda   }{3}   -
\bar{a}\,k_a - \bar{b}\,k_b .
\end{equation}
In what follows we will take the case $k_a=0=k_b$.

%%%%%%%%%%%%%%%%%%%%%%%%%%%%%%%%%%%%%%
\subsection{Non-Perturbative Solutions: Numerical Analysis and Late Times}
%%%%%%%%%%%%%%%%%%%%%%%%%%%%%%%%%%%%%%

In order to  extract qualitative physical information  from the model,
in the  present section  we will  perform a study  of the  behavior of
$\ba(\bt)$  and $\bbb(\bt)$  for  different regimes  of the  parameter
$\kappa$ and also for different values of $\lambda$.

For the non-perturbative  case, it is much better to  consider the set
of Hamiltonian equations (\ref{fo1}) to (\ref{fo4}) with  $k_a  =  0  =  k_b$.
Initial  conditions  for  the  system  are
$\ba(0)=r_a,  \bbb(0)  =r_b$,  and   for  numerical  solutions  we  use
$r_a=1=r_b$.  This  symmetric condition  just  encodes  the fact  that
patches of the universe are distinguished  at initial time only due to
the content of cosmological constant. Note also that these initial conditions
translate to the functions $a(t),b(t)$ as $a(0) = (\Lambda_a^2 + \Lambda_b^2)^{-1/4} =
b(0)$ so that the case $\Lambda_a=0=\Lambda_b$ is not included in the
rest of the discussion.

On the other hand, the initial  conditions for the momenta $\pi_a$ and
$\pi_b$ are  restricted by  the constraint  (\ref{constdimless}) which
can be written also as
\begin{equation}
\frac{\bpa^2}{\ba} + \frac{\bpb^2}{\bbb}  = \frac{\cos\lambda}{3}\ba^3 +
\frac{\sin\lambda}{3}\bbb^3.
\end{equation}

For the numerical solutions we choose
 \eqb
 \pi_a (0) &=& r_a^2  \sqrt{\frac{\cos \lambda}{3}}, ~~~~~~~~~~~ \pi_b
 (0) = r_b^2 \sqrt{\frac{\sin \lambda}{3}}.
 \label{f3}
 \eqf
This symmetric choice of initial conditions fulfills  the constraint (\ref{constdimless}) at $t=0$ and, since the Hamiltonian  is  preserved
 during  the evolution of  system, it is  satisfied at any time.

In our  numerical study we  are interested in the behavior of four quantities in the patch described by $a$. Namely, the scale  factor $a$, the  velocity of  the
 expansions $\dot{a}$, the  Hubble parameter $H_a$, and the  deceleration parameter $q_a$.  The
 last two are defined as follow
 \begin{equation}
 \label{hq}
 H_{a}   =   \frac{\dot{a}}{a}   =\mu\frac{\dot{\ba}}{\ba}\equiv   \mu
 \bar{H}_\ba, ~~~~~~~~~ q_a = -\frac{\ddot{a}}{\dot{a}^2}a =
 -\frac{\ddot{\ba}}{\dot{\ba}^2}\ba,
 \end{equation}
where time derivatives of quantities with a bar are taken with
respect to $\bt$. Similar  definitions hold for  the scale  factor $\bbb$.  Note finally  that  $q_a$ is
independent of the scale $\mu$.

It is  interesting to explore  the cases $\Lambda_a \ll \Lambda_b$ and
$\Lambda_a  \sim  \Lambda_b$ separately. The  case  $\Lambda_a  \gg \Lambda_b$  is
contained  in  the   first,  due  to  the  symmetry  $a   \to  b$  and
$\kappa \to -\kappa$  . Then we will study the  quantities of interest
in such limits for different values of $\kappa$

%%%%%%%%%%%%%%%%%%%%%%%%%%%%%%
\subsubsection{The case $\Lambda_a \ll \Lambda_b$}
\begin{figure}[tb]
\centering
\begin{minipage}[c]{0.45\textwidth}
\subfigure[\label{1a}]{
\includegraphics[scale=0.6]{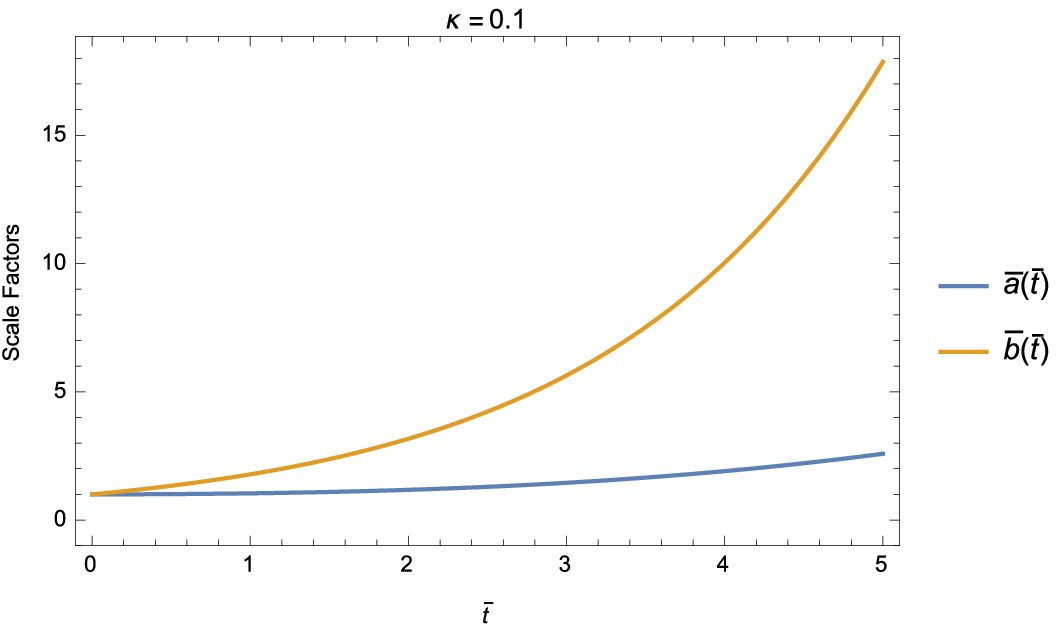}
\label{fig-abk01}
}
\end{minipage}
\begin{minipage}[c]{0.45\textwidth}
\subfigure[\label{1b}]{
\includegraphics[scale=0.6]{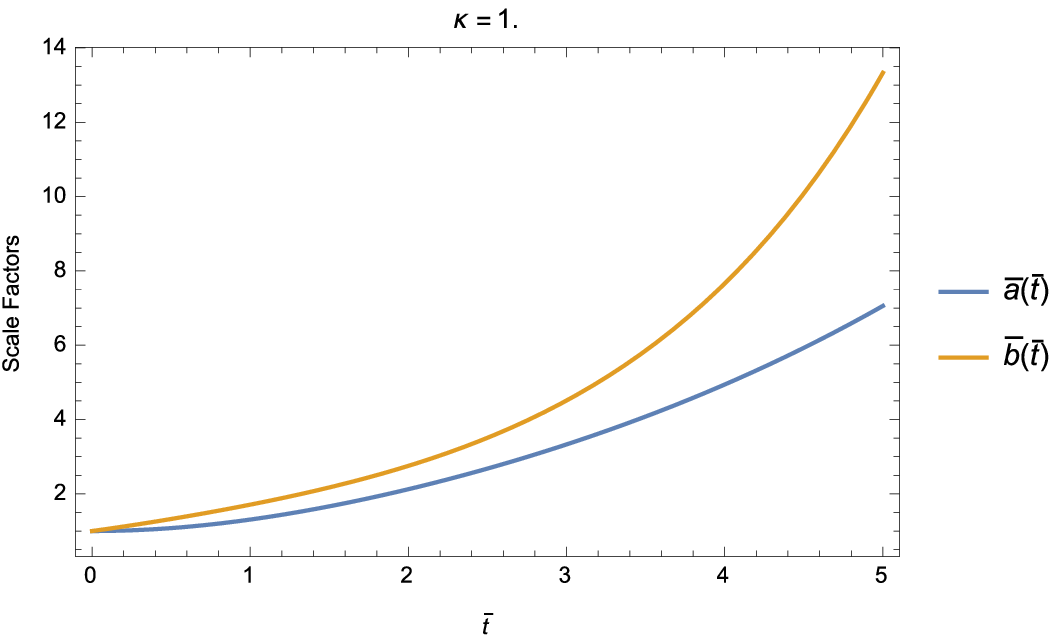}
\label{fig-abk1}
}
\end{minipage}
\begin{minipage}[c]{0.45\textwidth}
\subfigure[\label{1c}]{
\includegraphics[scale=0.6]{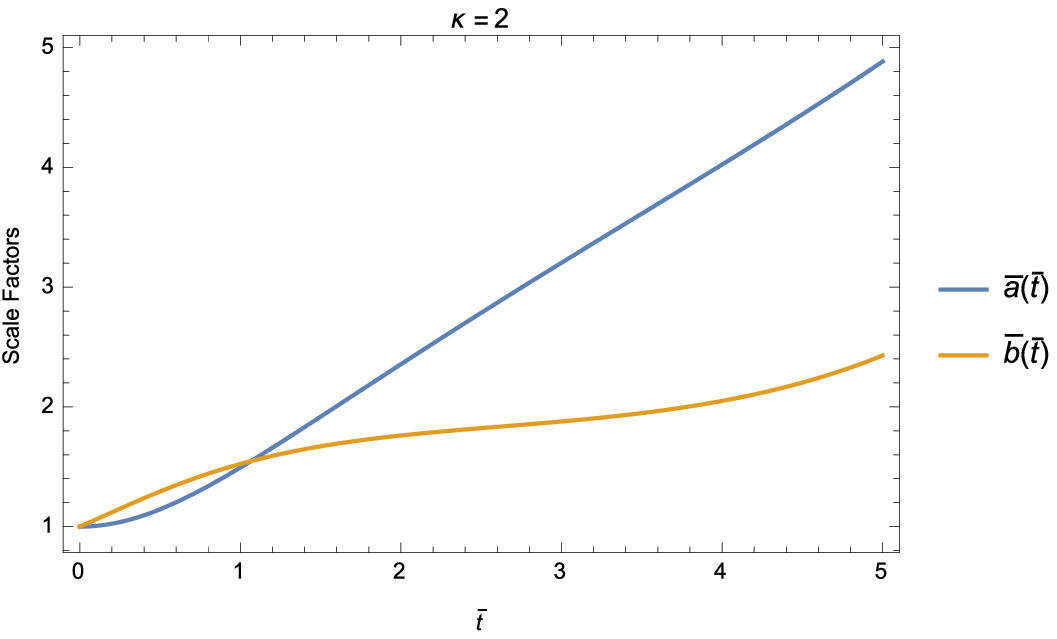}
\label{fig-abk2}
}
\end{minipage}
\begin{minipage}[c]{0.45\textwidth}
\subfigure[\label{1d}]{
\includegraphics[scale=0.6]{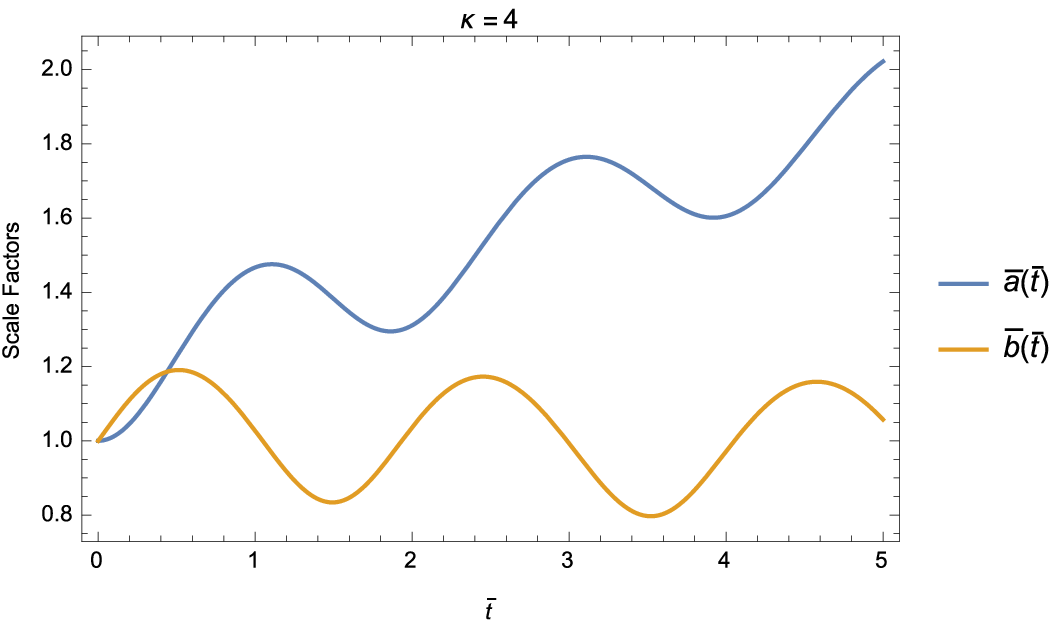}
\label{fig-abk4}
}
\end{minipage}
\caption{\small{ Scale  factors for  different values of  $\kappa$ and
    for  $\Lambda_a  = \mu^2 \sin\epsilon$, $\Lambda_b = \mu^2 \cos\epsilon$, $\epsilon=10^{-4}$.}}
\label{fig:fig1}
\end{figure}

%%%%%%%%%%%%%%%%%%%%%%%%%%%%%%
We          first     examine  the       case   $\Lambda_a \ll \Lambda_b$,    which we illustrate by
$\Lambda_a  = \mu^2 \sin\epsilon \approx \epsilon  \mu^2 $, $\Lambda_b = \mu^2 \cos\epsilon\approx (1-\epsilon^2/2)  \mu^2 $, $\epsilon=10^{-4}$.   The  behavior  of the  scale  factors $a$ and $b$  for
$\kappa=0.1$,  $\kappa=1$,  $\kappa=2$  and $\kappa=4$  is  shown  in
Figure \ref{fig:fig1}.

We observe here that as $\kappa$ increases, the scale factors $\ba$ and $\bbb$
start an exponential growth. Indeed, while in panels \ref{1a} and \ref{1b} we see that  $\bbb(\bt)
> \ba(\bt)$, the situation is reversed in panels \ref{1c} and \ref{1d} for $\bt \agt 1$.

On the other hand, as   $\kappa$ increases further,  the system
exhibits   a quasi   periodic   behavior  as   can  be   checked  in panels
\ref{1c} and \ref{1d} in the same figure.  The oscillation  pattern also shows  how  $\ba$  grows  at the
expenses of $\bbb$.

%%%%%%%%%%%%%%%%%%%%%%%%%%%%%%%%
%%%%%% F I G U R E 1 %%%%%%%%%%%%
%%%%%%%%%%%%%%%%%%%%%%%%%%%%%%%%

%%%%%%%%%%%%%%%%%%%%%%%%%%
%%%%%%%%%%%%%%%%%%%%%%%%%%%

\begin{figure}[btp]
\centering
\begin{minipage}[c]{0.45\textwidth}
\subfigure[\label{n1a}]{
\includegraphics[scale=0.6]{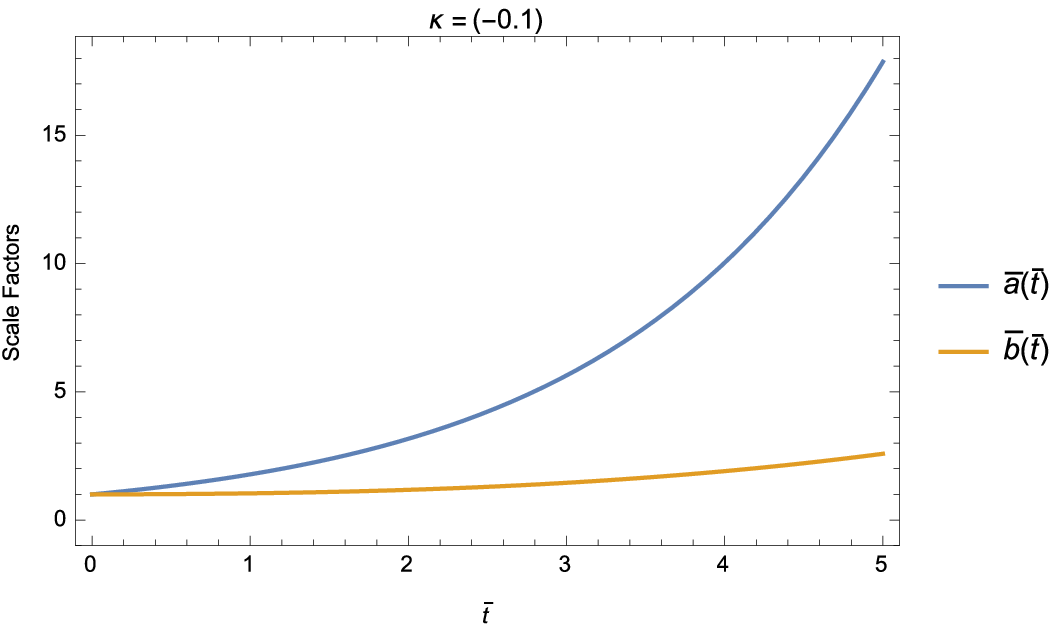}
}
\end{minipage}
\begin{minipage}[c]{0.45\textwidth}
\subfigure[\label{n1b}]{
\includegraphics[scale=0.6]{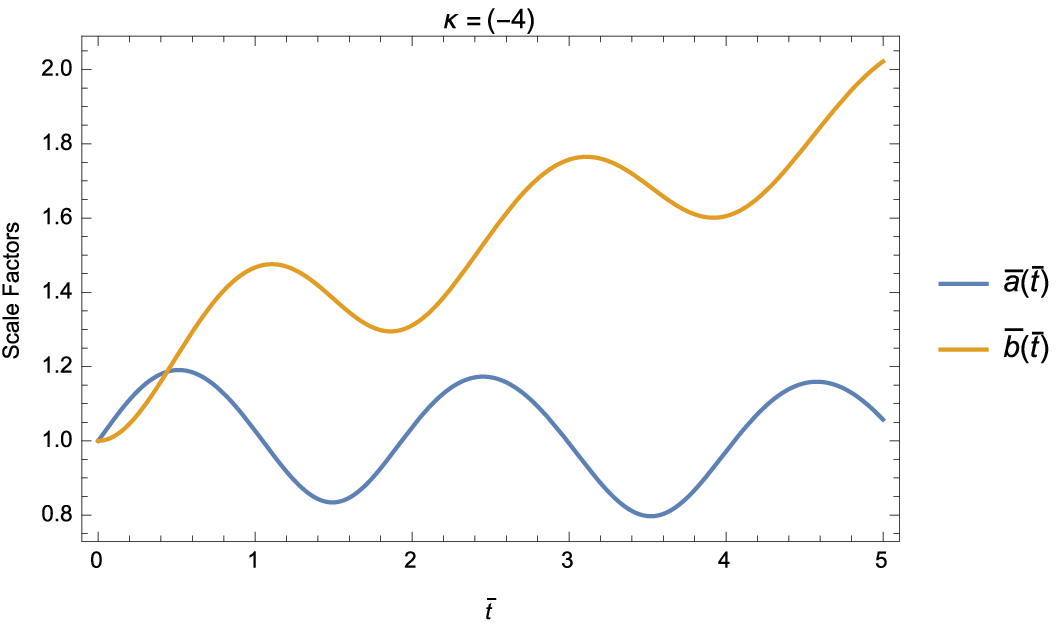}
\label{fig-abk4}
}
\end{minipage}
\caption{\small{ Scale  factors for  negative values of  $\kappa$ and
    for $\Lambda_b  = \mu^2 \sin\epsilon \approx \epsilon \mu^2$, $\Lambda_a = \mu^2 \cos\epsilon \approx (1-\epsilon^2/2) \mu^2$, $\epsilon=10^{-4}$. The plots are the same as in Figure \ref{1a} and \ref{1d} after changing $\ba \leftrightarrow \bbb$.}}
\label{fig:nfig1}
\end{figure}

Figure \ref{fig:nfig1} shows the behavior of the  scale
factors for negative $\kappa$ and $\Lambda_a \leftrightarrow \Lambda_b$.
We show only the cases $\kappa = -0.1$ and
$\kappa=-4$.
We verify here  our statement that the case $\Lambda_b \ll \Lambda_a$ is already
contained in the present discussion.

%%%%%%%%%%%%%%%%%%%%%%%%%%%%%%%%
%%%%%% F I G U R E 1 NEGATIVES%%%%%%%%%%%%
%%%%%%%%%%%%%%%%%%%%%%%%%%%%%%%%
From here on we plot only the values of $\kappa$ that exhibit the main features
we would like to highlight.

%%%%%%%%%%%%%%%%%%%%%%%%%%
%%%%%%%%%%%%%%%%%%%%%%%%%%%

\begin{figure}[btp]
\centering
\begin{minipage}[c]{0.45\textwidth}
\subfigure[\label{hub1a}]{
\includegraphics[scale=0.63]{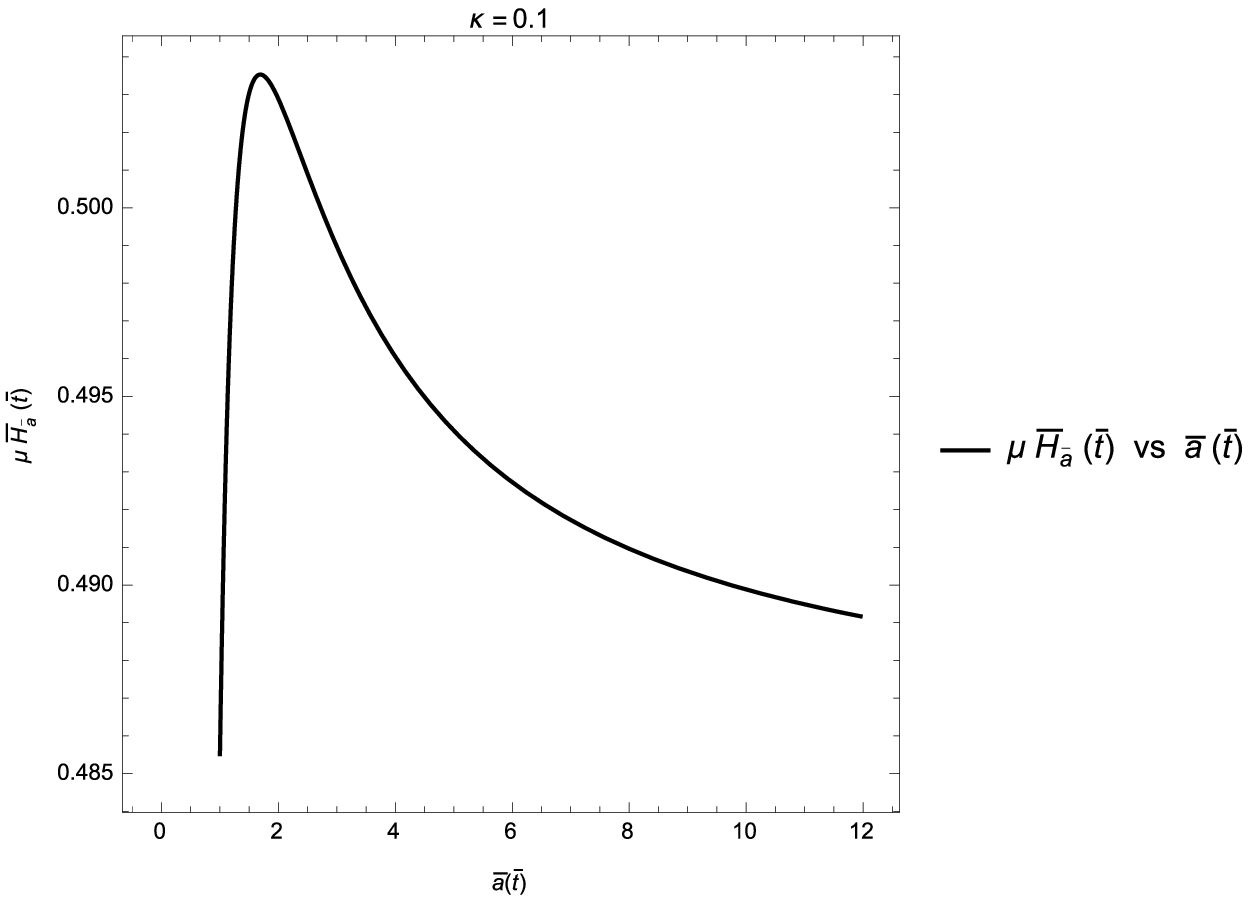}}
\end{minipage}
\begin{minipage}[c]{0.45\textwidth}
\subfigure[\label{hub1b}]{
\includegraphics[scale=0.63]{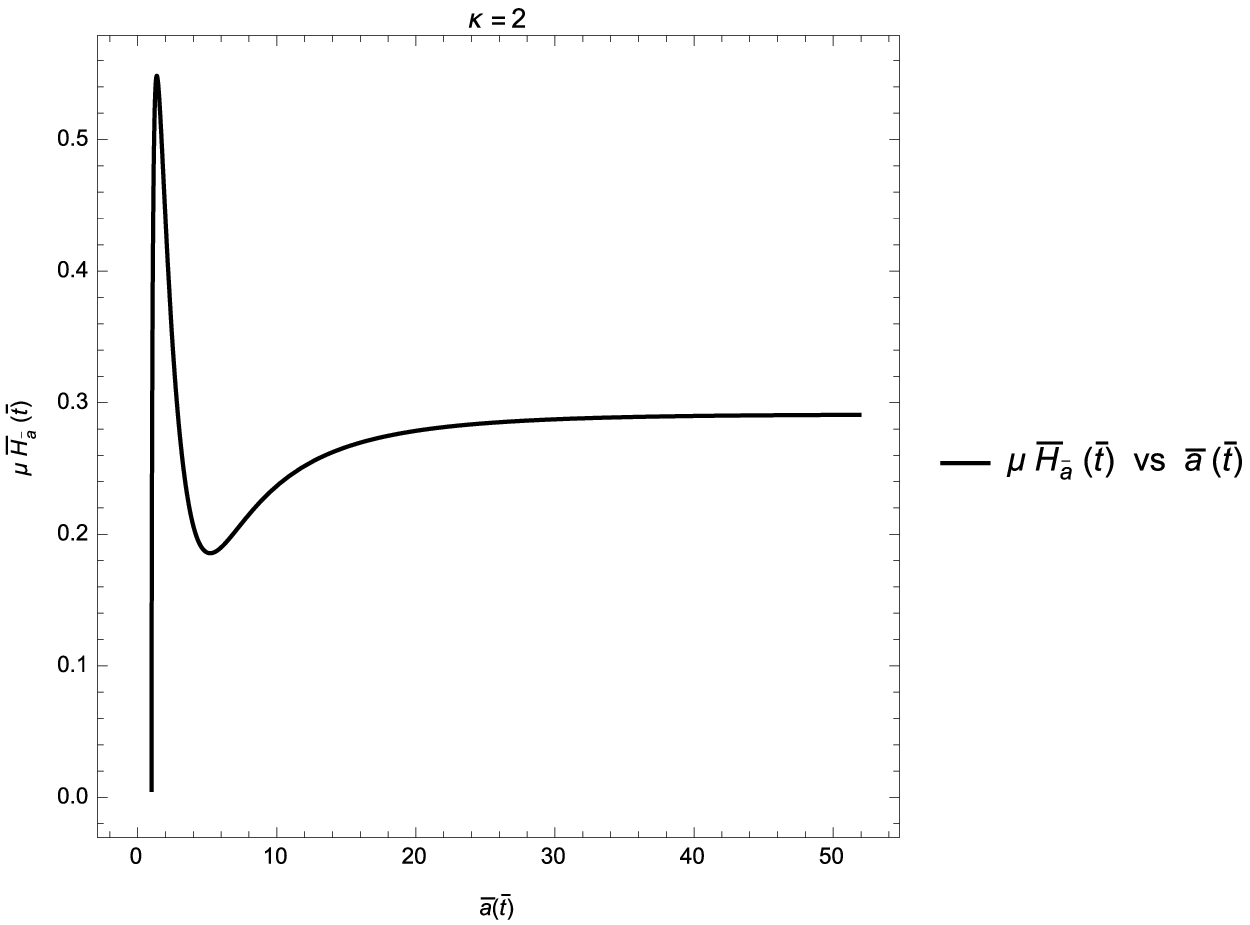}
}
\end{minipage}
\begin{minipage}[c]{0.45\textwidth}
\subfigure[\label{hub1c}]{
\includegraphics[scale=0.63]{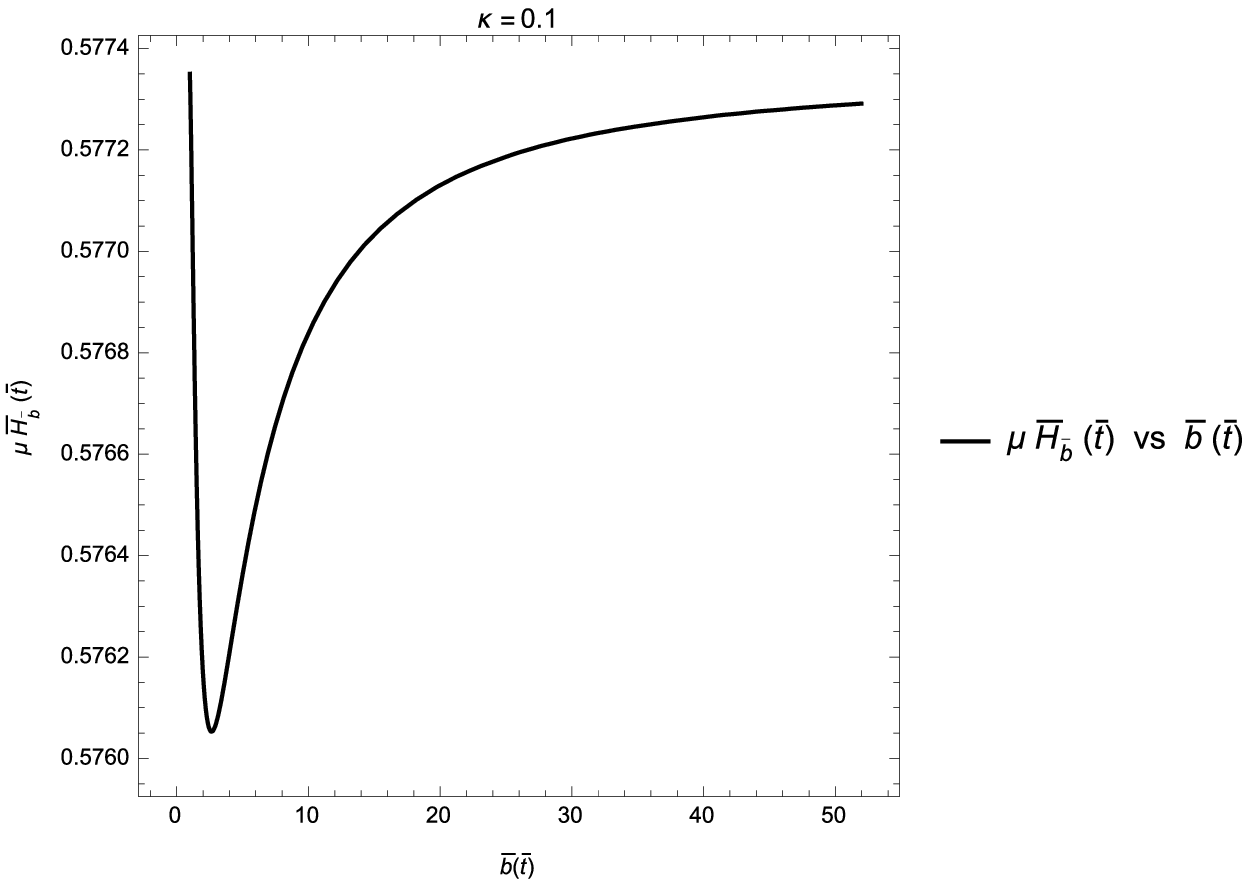}
}
\end{minipage}
\begin{minipage}[c]{0.45\textwidth}
\subfigure[\label{hub1d}]{
\includegraphics[scale=0.63]{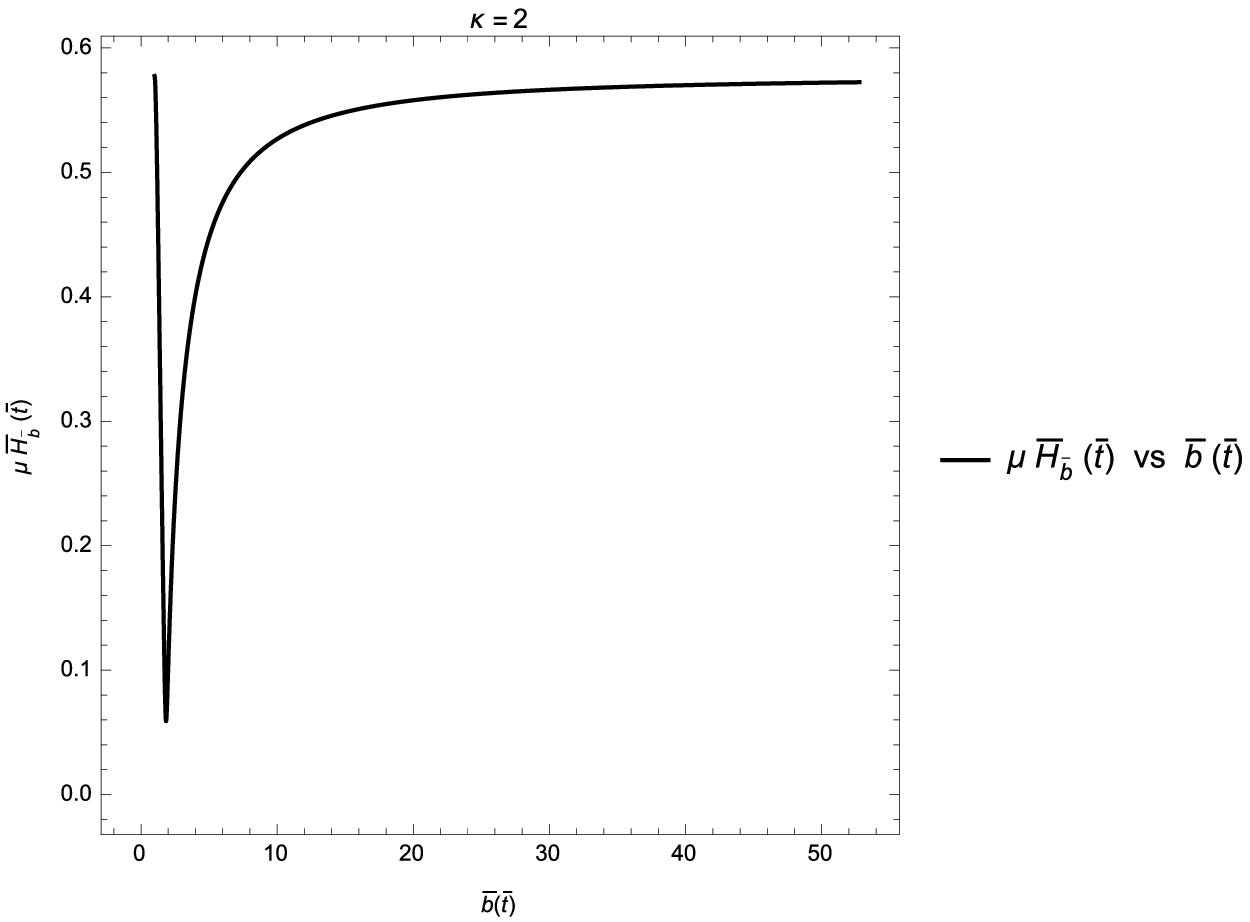}
}
\end{minipage}
\caption{\small{ Parametric plot of Hubble parameter as a function of the scale factor for different values of $\kappa$ and for $\Lambda_a \ll \Lambda_b$. On the horizontal axes we can read the value of the scale factor at different times.}}
\label{fig:fighub1}
\end{figure}

The Hubble parameter  $H_a$ defined in  (\ref{hq})  also has
an interesting behavior.  Figure \ref{fig:fighub1} shows the evolution of
$H_a$ and $H_b$ as  a function of  their respective scale factor. For the scale factor
$\ba$ we observe a different behavior of the Hubble parameter for
$\kappa = 0.1$ in panel \ref{hub1a} compared with  the case $\kappa = 2$ in panel
\ref{hub1b}. Similarly for the scale factor  $\bbb$ shown in panels \ref{hub1c} and
\ref{hub1d} for same values of $\kappa$.

It is interesting again to note  the {\it complementary} behavior of $\ba$
and $\bbb$, in the sense that the increase in one of the Hubble parameters
is accompanied by the decrease of the Hubble parameter of the other scale factor.

Indeed, for $\lambda=\pi/2$  and $\kappa=0$, $\bar{H}_\ba = 0$ since the
solution  of  the   equations  of  motion  in  such   a  situation  is
$\ba(\bt) =  $ constant. The Hubble parameter for $\bbb (\bt)$ in such case
is a non zero constant ($\mu^2\bar{H}_\bbb = 3^{-1/2} = 0.57735$). Panel \ref{hub1a}
shows how the Hubble parameter for $\ba$ going from $0$ to $\approx 0.31$ in $\Delta \ba \approx 2$
while in panel \ref{hub1c} the Hubble parameter for $\bbb$ diminishes from $\approx 0.57735$ to
$\approx 0.576005$ in $\Delta\bbb\approx 2$.

This kind of complementarity is also observed in panels \ref{hub1c} and
\ref{hub1d}, where the increase-decrease process occurs for $\Delta\ba\approx 10\approx
\Delta\bbb$.

 %%%%%%%%%%%%%%%%%%%%%%%%%%%%%%%%%
%%%%%% F I G U R E 3-V2%%%%%%%%%%%%
%%%%%%%%%%%%%%%%%%%%%%%%%%%%%%%%

%%%%%%%%%%%%%%%%%%%%%%%%
%%%%%%%%%%%%%%%%%%%%%%%%%

\begin{figure}[btp]
\centering
\begin{minipage}[c]{0.45\textwidth}
\subfigure[\label{velac1a}]{
\includegraphics[scale=0.66]{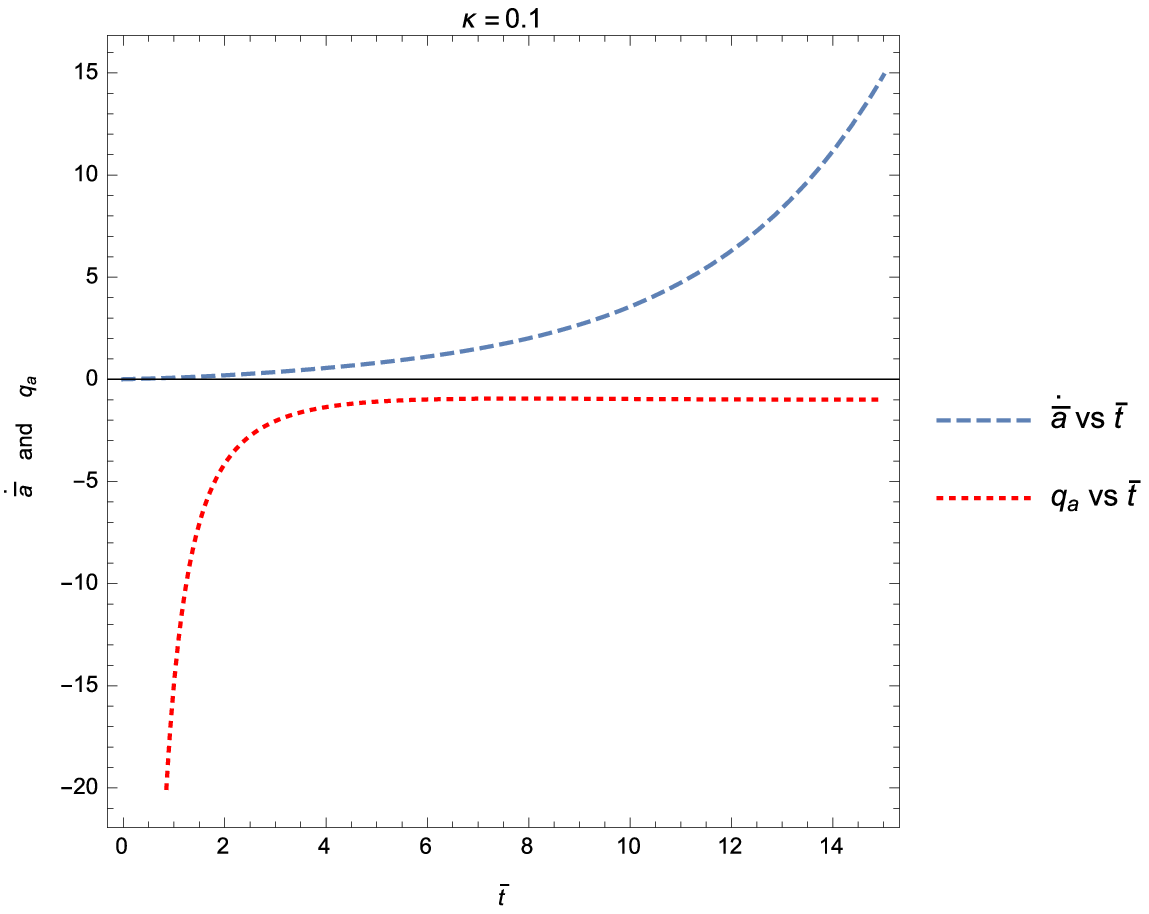}
}
\end{minipage}
\begin{minipage}[c]{0.45\textwidth}
\subfigure[\label{velac1b}]{
\includegraphics[scale=0.66]{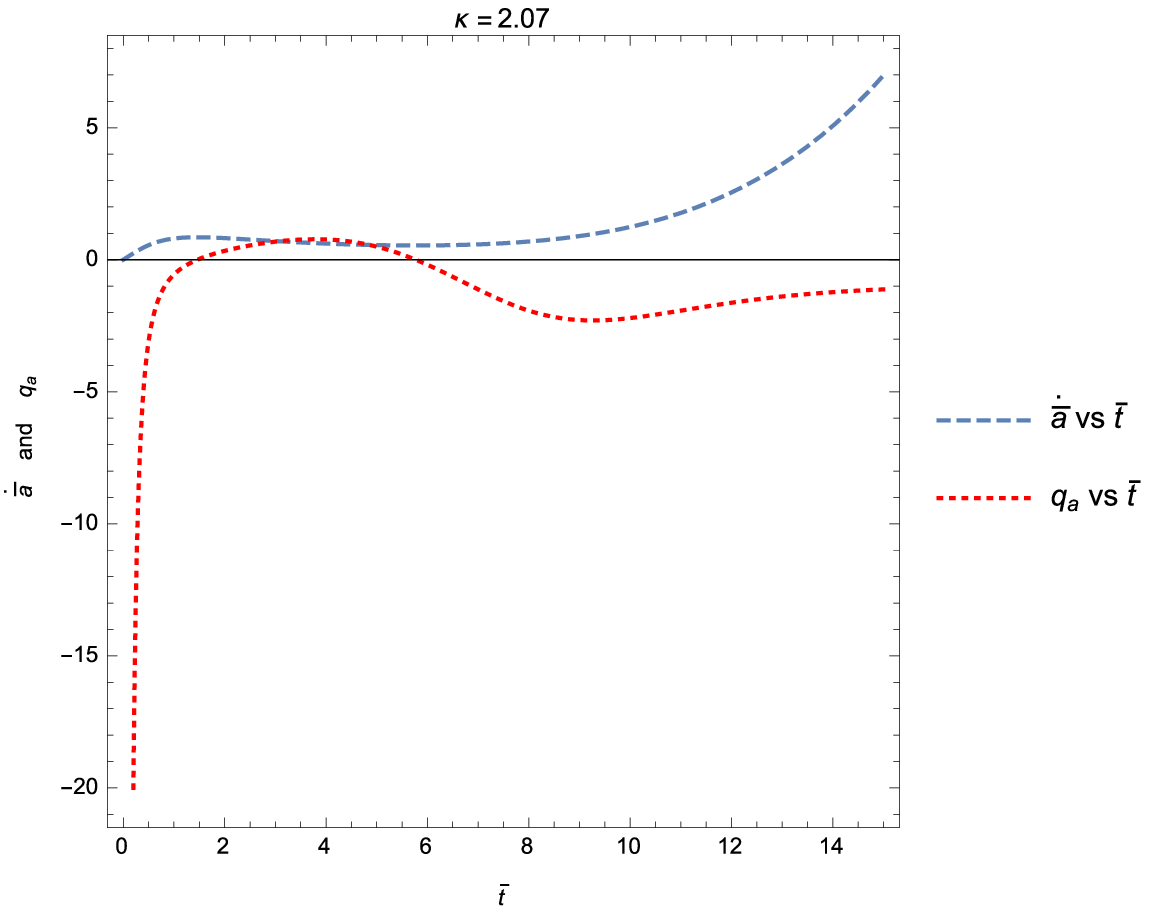}
}
\end{minipage}
\begin{minipage}[c]{0.45\textwidth}
\subfigure[\label{velac1c}]{
\includegraphics[scale=0.66]{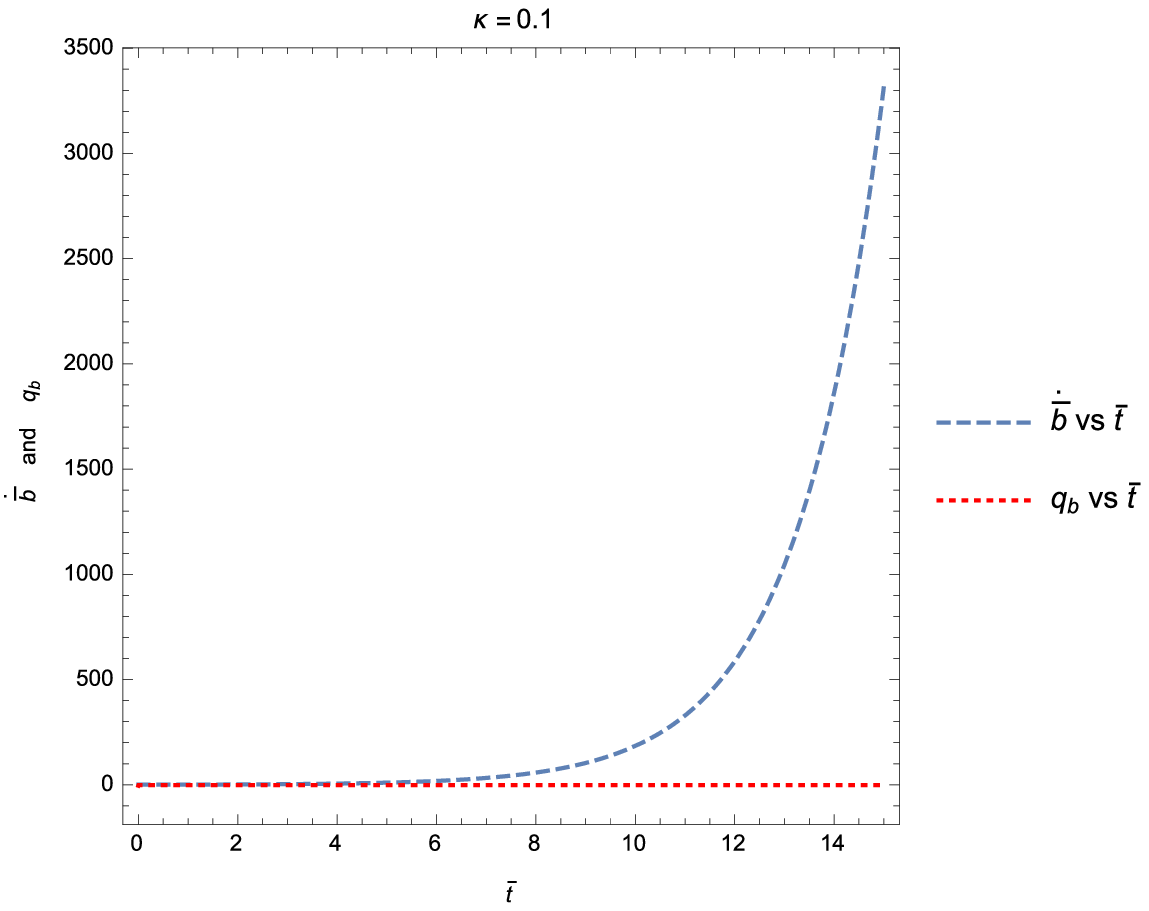}
}
\end{minipage}
\begin{minipage}[c]{0.45\textwidth}
\subfigure[\label{velac1d}]{
\includegraphics[scale=0.66]{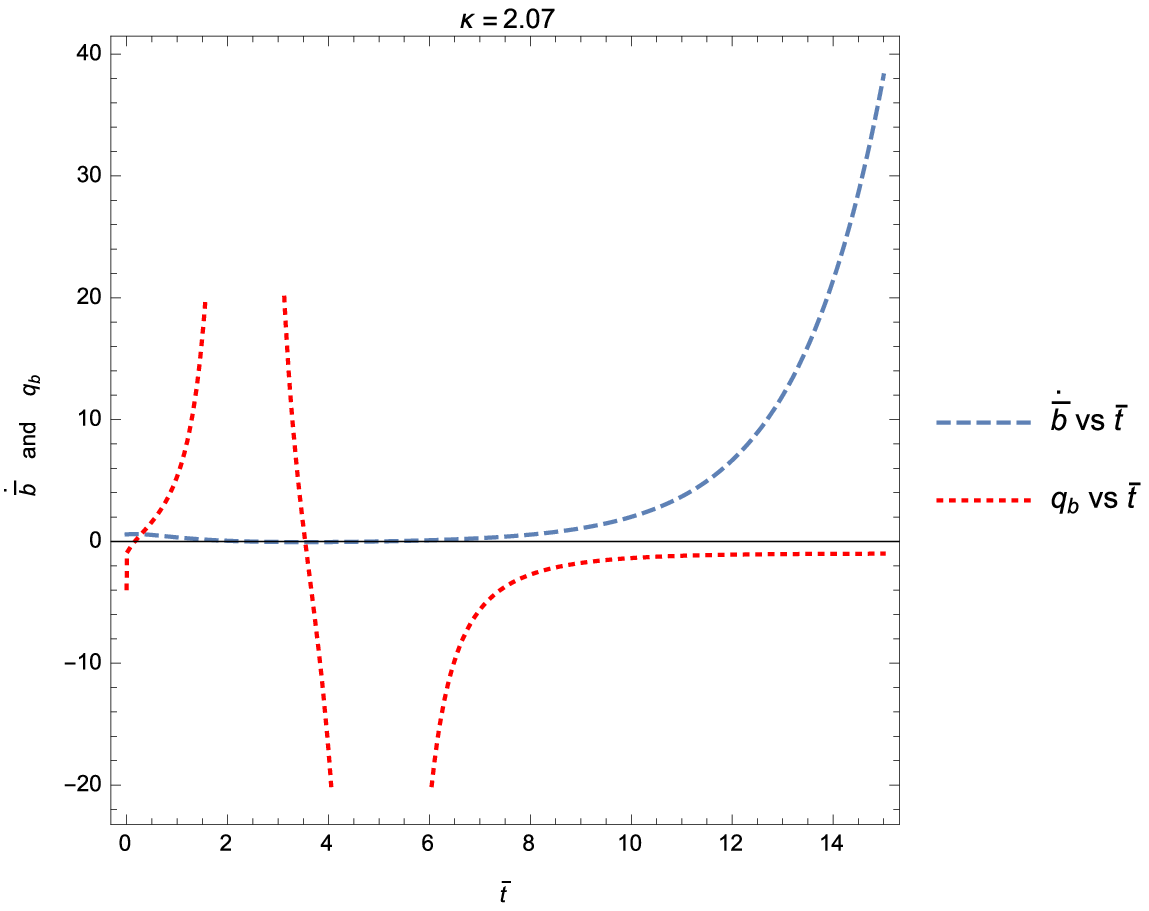}
}
\end{minipage}
\caption{\small{ Velocity  expansion and deceleration parameters for
 the scale factors $\ba$ and $\bbb$ as functions of $\bt$
 in the regime  $\Lambda_a \ll \Lambda_b$ for two different values of
 $\kappa$. The panels on the right show imprints of the quasi periodic behavior of
 the scale factors.}}
\label{fig:figavelac1}
\end{figure}

Figure \ref{fig:figavelac1} shows velocities and deceleration
parameters of  $\ba$ and $\bbb$ for two different values of
$\kappa$. For $\kappa=2.07$ we observe the imprints of
the quasi periodic behavior of the scale factor. In panel \ref{velac1b}
we appreciate an increasing deceleration parameter, which starts to  decrease
at $\bt\sim 6$ and for $\bt \agt 12$ starts to stabilize to zero.

The  increase of the  velocity expansion of $\ba$ due to  the interaction
with  $\bbb$  can be  appreciated  as  $\kappa$ increases.   While  the
velocity  of  the   $\ba$ patch increases,  the velocity  of  the  $\bbb$ patch
decreases.

%%%%%%%%%%%%%%%%%%%%%%%%%%%%%%%%
%%%%%% F I G U R E 4-V2 %%%%%%%%%%%%
%%%%%%%%%%%%%%%%%%%%%%%%%%%%%%%%

%%%%%%%%%%%%%%%%%%
%%%%%%%%%%%%%%%%%%
\subsubsection{The case $\Lambda_a \sim \Lambda_b$}

\begin{figure}[btp]
\centering
\begin{minipage}[c]{0.45\textwidth}
\subfigure[\label{absima}]{
\includegraphics[scale=0.5]{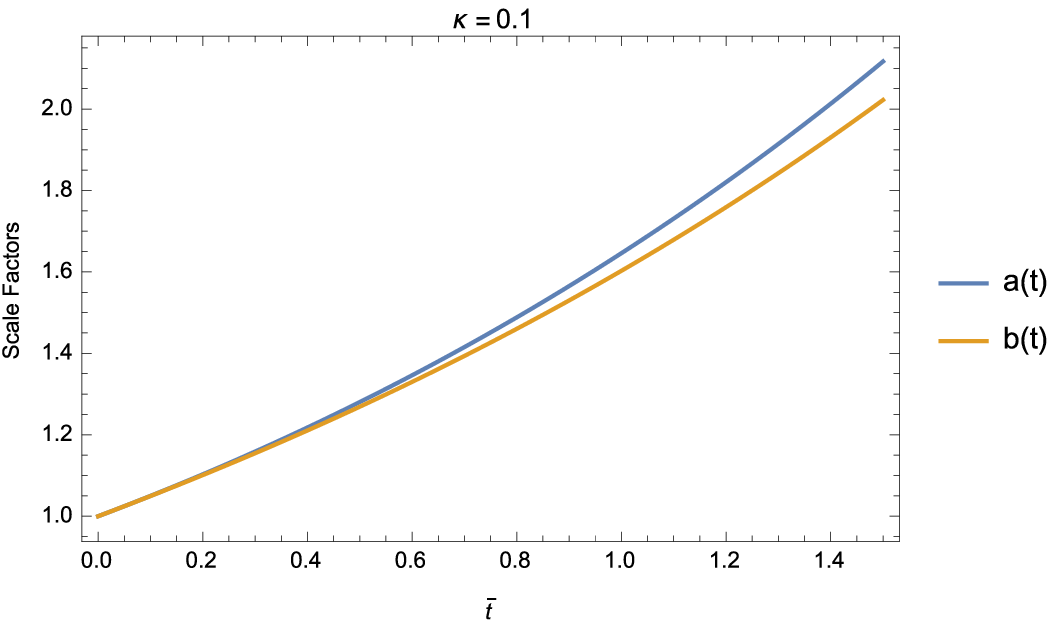}
}
\end{minipage}
\begin{minipage}[c]{0.45\textwidth}
\subfigure[\label{absimb}]{
\includegraphics[scale=0.5]{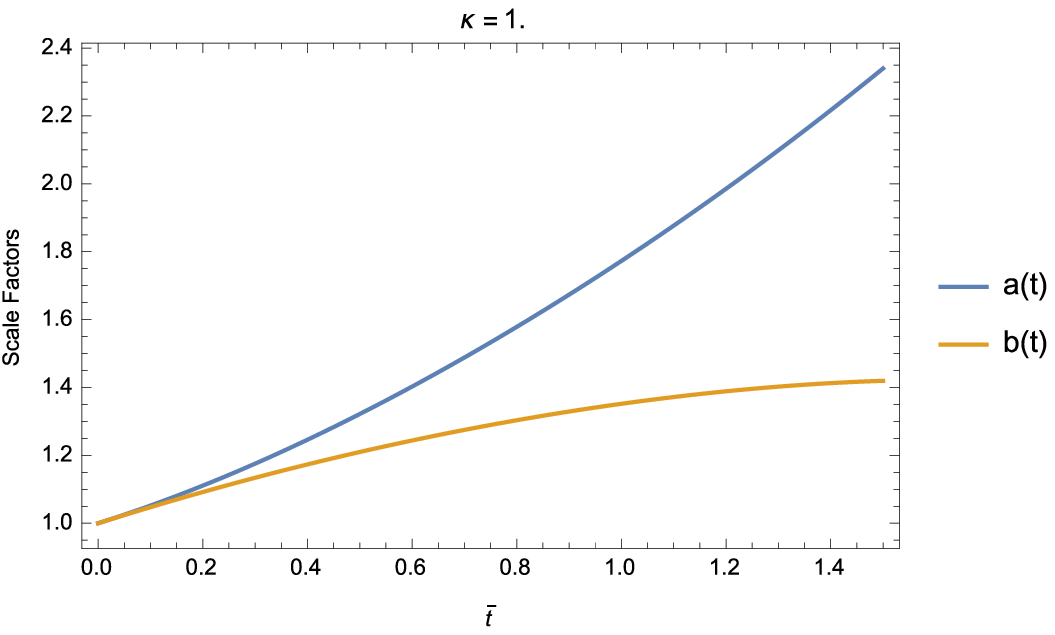}
\label{fig- absimk1}
}
\end{minipage}
\begin{minipage}[c]{0.45\textwidth}
\subfigure[\label{absimc}]{
\includegraphics[scale=0.5]{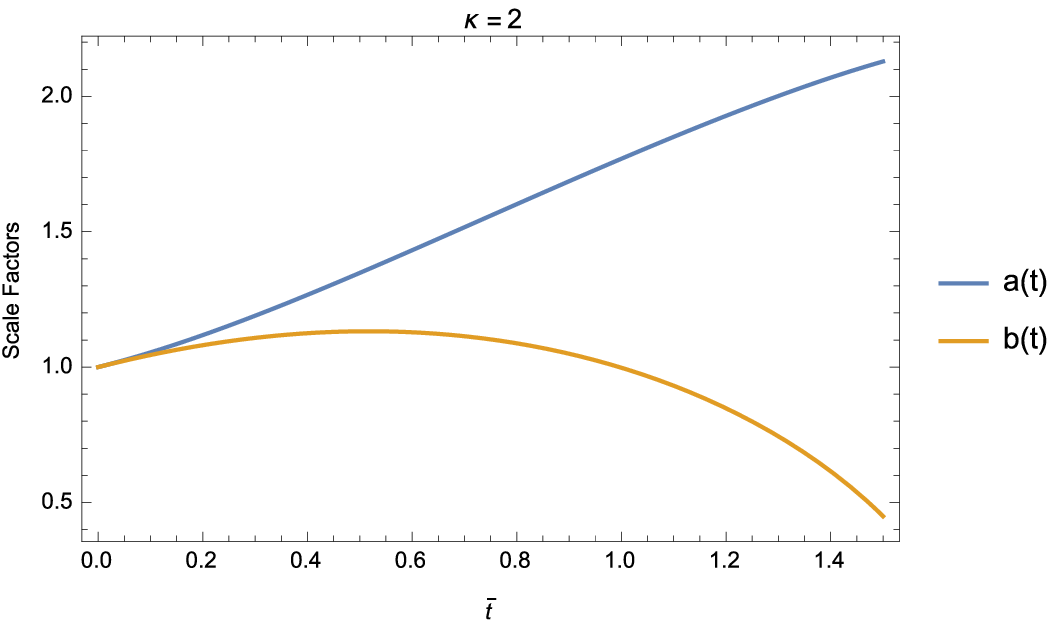}
\label{fig-absimk2}
}
\end{minipage}
\begin{minipage}[c]{0.45\textwidth}
\subfigure[\label{absimd}]{
\includegraphics[scale=0.5]{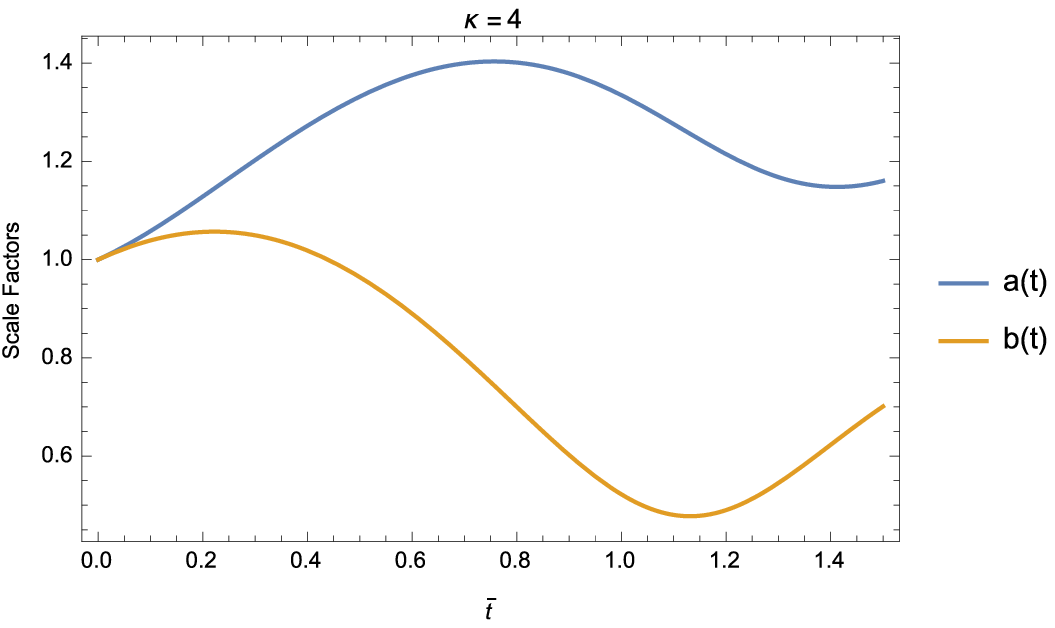}
\label{fig-absimk4}
}
\end{minipage}
\caption{\small{ Scale  factors $\ba$ and $\bbb$ for  different values of  $\kappa$ and
    for $\Lambda_a \approx \Lambda_b$. }}
\label{fig:figabsim}
\end{figure}

%%%%%%%%%%%%%%%%%%
%%%%%%%%%%%%%%%%%%
Let us consider   the  case   $\Lambda_a\sim\Lambda_b$. For illustration, we have taken  $\lambda =  (\pi/4) - \epsilon$ with $\epsilon=10^{-4}$.  The scale  factor  behavior can  be appreciated  in
Figure \ref{fig:figabsim}. We note that  the effect of $\kappa$  is to
increase    the   exponential    growth    of    the   scale    factor
$\ba$.    The        cosmological            constants           satisfy
$\Lambda_a \approx  (1 +\epsilon) \Lambda_b$  and  we  expect the scale
factor $\ba$ to increase faster that  $\bbb$. This is what panel
\ref{absima} shows.  In  the rest of the  panels we see how  the faster increase of $\ba$
  becomes more pronounced as $\kappa$ grows. The quasi periodic behavior of $\ba$ and $\bbb$
can be appreciated in panels \ref{absimc} and \ref{absimd}.

%%%%%%%%%%%%%%%%%%%%%%%%%%%%%%%%
%%%%%% F I G U R E 5 -VER2 %%%%%%%%%%%%
%%%%%%%%%%%%%%%%%%%%%%%%%%%%%%%%
%%%%%%%%%%%%%%%%%%%%%%%%%%
%%%%%%%%%%%%%%%%%%%%%%%%%%

\begin{figure}[btp]
\centering
\begin{minipage}[c]{0.45\textwidth}
\subfigure[\label{hubsima}]{
\includegraphics[scale=0.61]{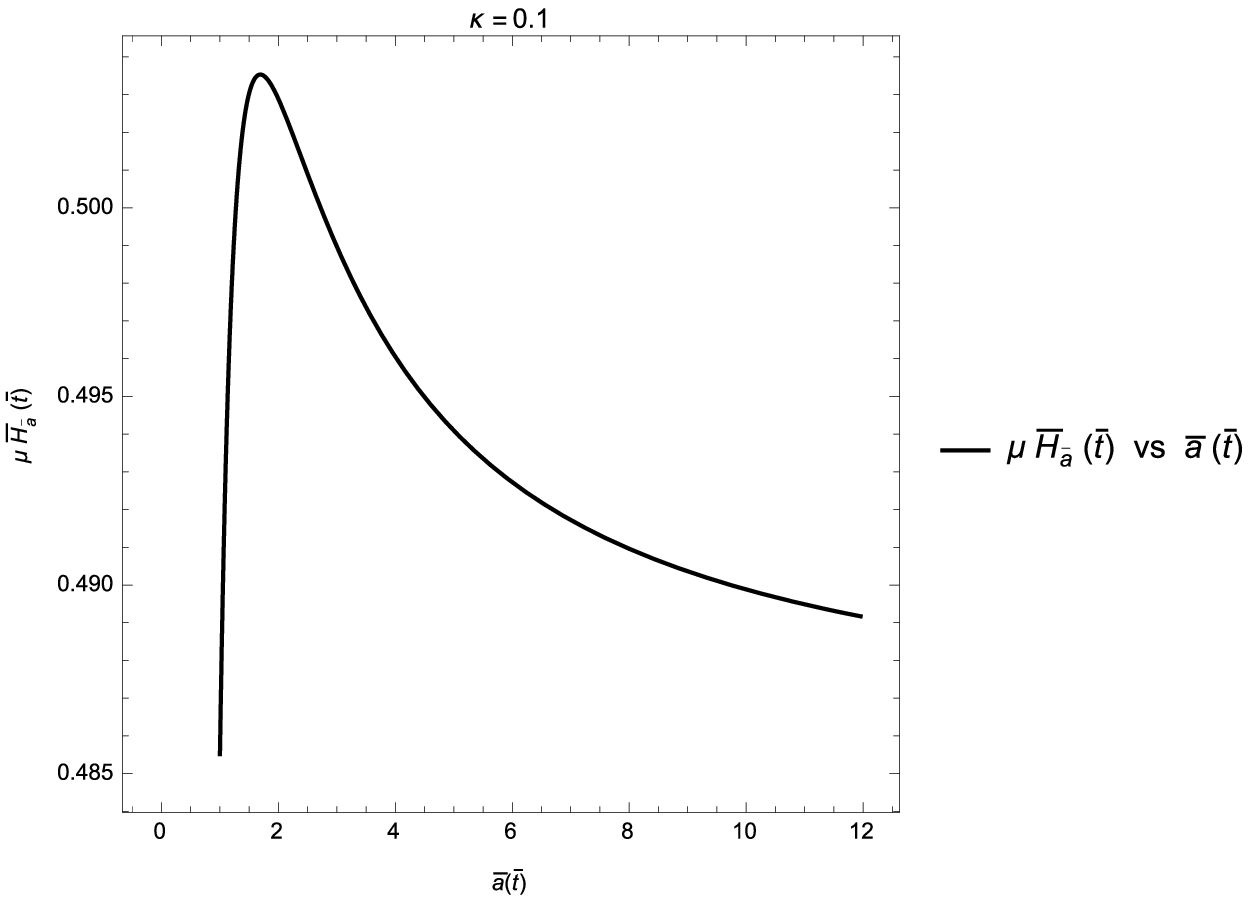}
\label{fig-hubsim01}
}
\end{minipage}
\begin{minipage}[c]{0.45\textwidth}
\subfigure[\label{hubsimb}]{
\includegraphics[scale=0.61]{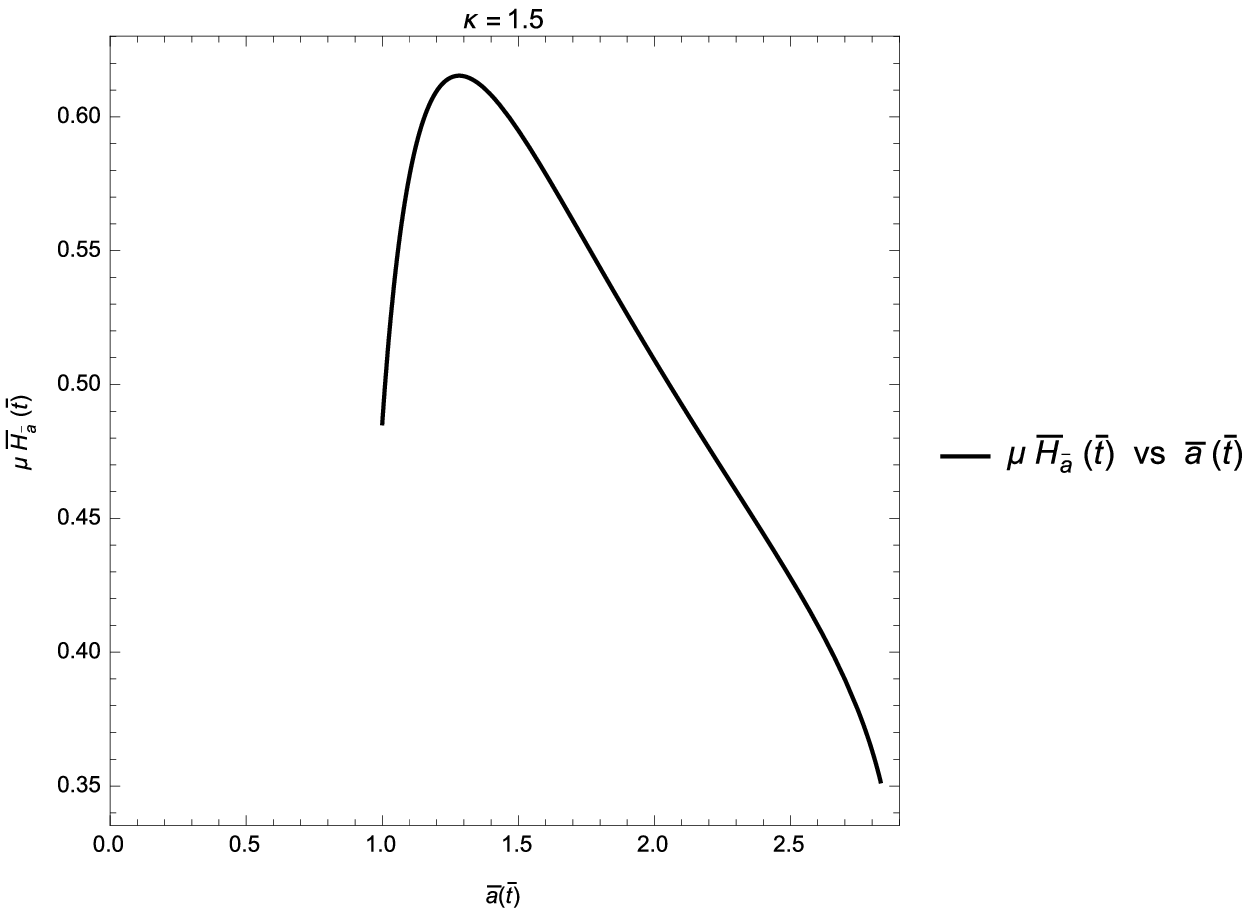}
\label{fig- hubsimk1}
}
\end{minipage}
\begin{minipage}[c]{0.45\textwidth}
\subfigure[\label{hubsimc}]{
\includegraphics[scale=0.61]{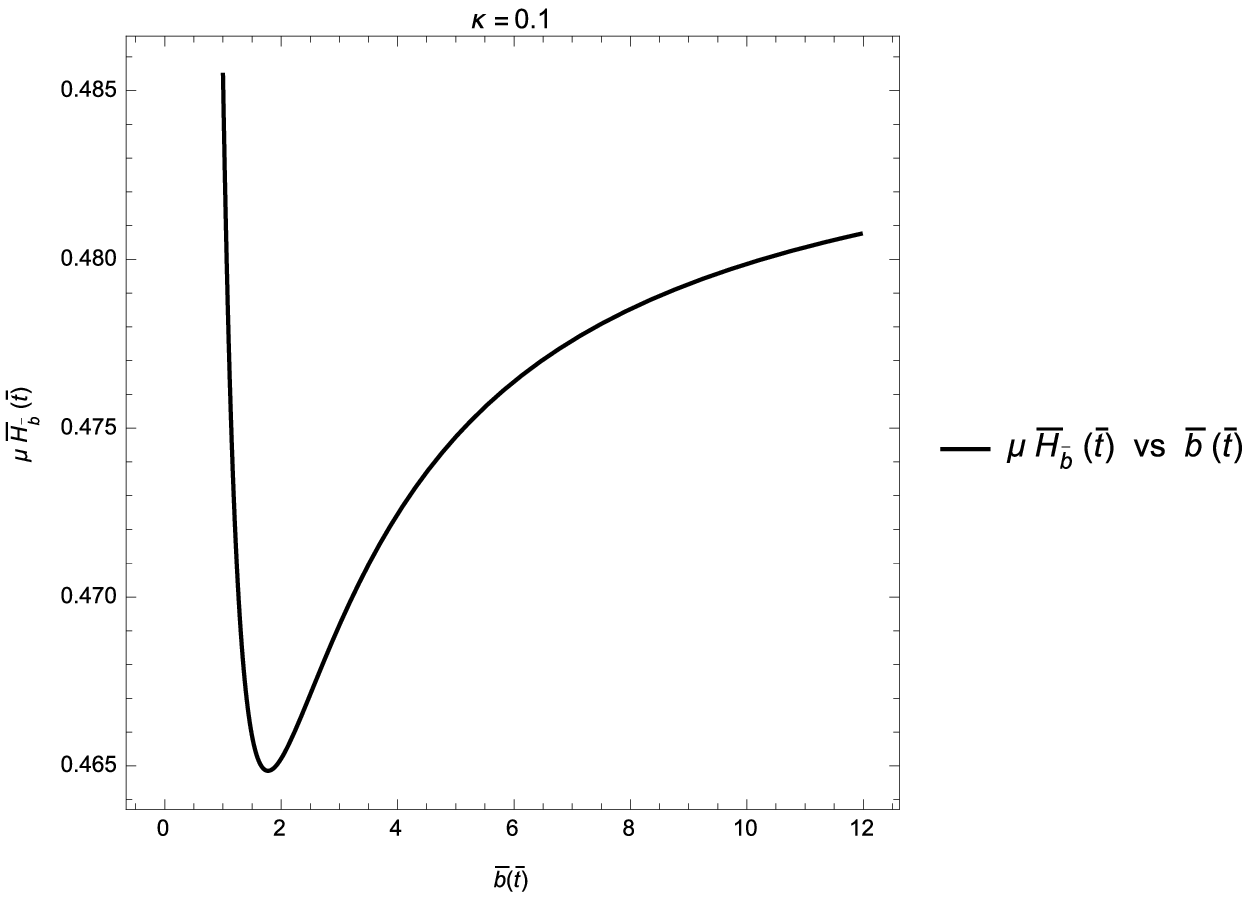}
\label{fig-hubsimk2}
}
\end{minipage}
\begin{minipage}[c]{0.45\textwidth}
\subfigure[\label{hubsimd}]{
\includegraphics[scale=0.61]{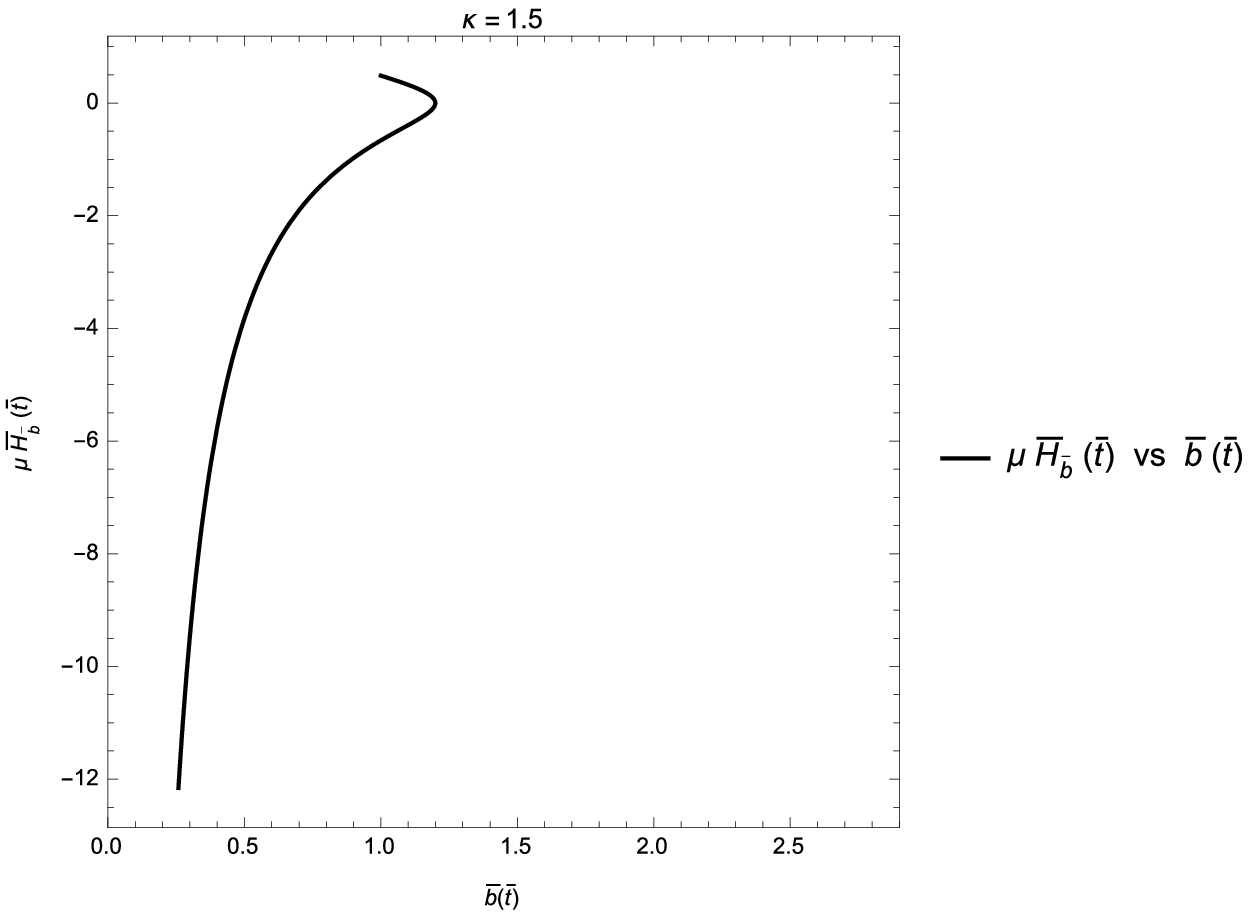}
\label{fig-hubsimk4}
}
\end{minipage}
\caption{\small{  Hubble parameters  vs  scale  factors for  different
    values of $\kappa$  and for $\Lambda_a \approx  \Lambda_b$.}}
\label{fig:fighubsim}
\end{figure}

The   Hubble  constant   behavior   can  be   appreciated  in   figure
\ref{fig:fighubsim}. Panel \ref{hubsima}  shows how the expansion  rate of $\ba$
increases, and  panel \ref{hubsimc} shows the decrease of  the expansion rate
of $\bbb$, both cases for the same value of $\kappa = 0.1$. This is in agreement
with the behavior of $\ba$ and $\bbb$ previously shown in panel \ref{absima}.
The situation  changes  as  $\kappa$ increases. The scale factor $\bbb$  decreases
 from its initial value $\bbb(0) = 1$ and $\ba$ starts to grow (see panels \ref{absimb} and
 \ref{absimc} in  Figure \ref{fig:figabsim}, and note that in this figure we take $\kappa=1.5$).  The expansion
  rate of $\ba$, however, is  greater than  the expansion rate  of $\bbb$, but  in any  case always
decreases.
%%%%%%%%%%%%%%%%%%%%%%%%%%%%%%%%
%%%%%% F I G U R E 6-VER2 %%%%%%%%%%%%
%%%%%%%%%%%%%%%%%%%%%%%%%%%%%%%%

%%%%%%%%%%%%%%%%%%%%%%%%%%%%%%%%
%%%%%%%%%%%%%%%%%%%%%%%%%%%%%%%%

\begin{figure}[btp]
\centering
\begin{minipage}[c]{0.45\textwidth}
\subfigure[\label{velaca}]{
\includegraphics[scale=0.65]{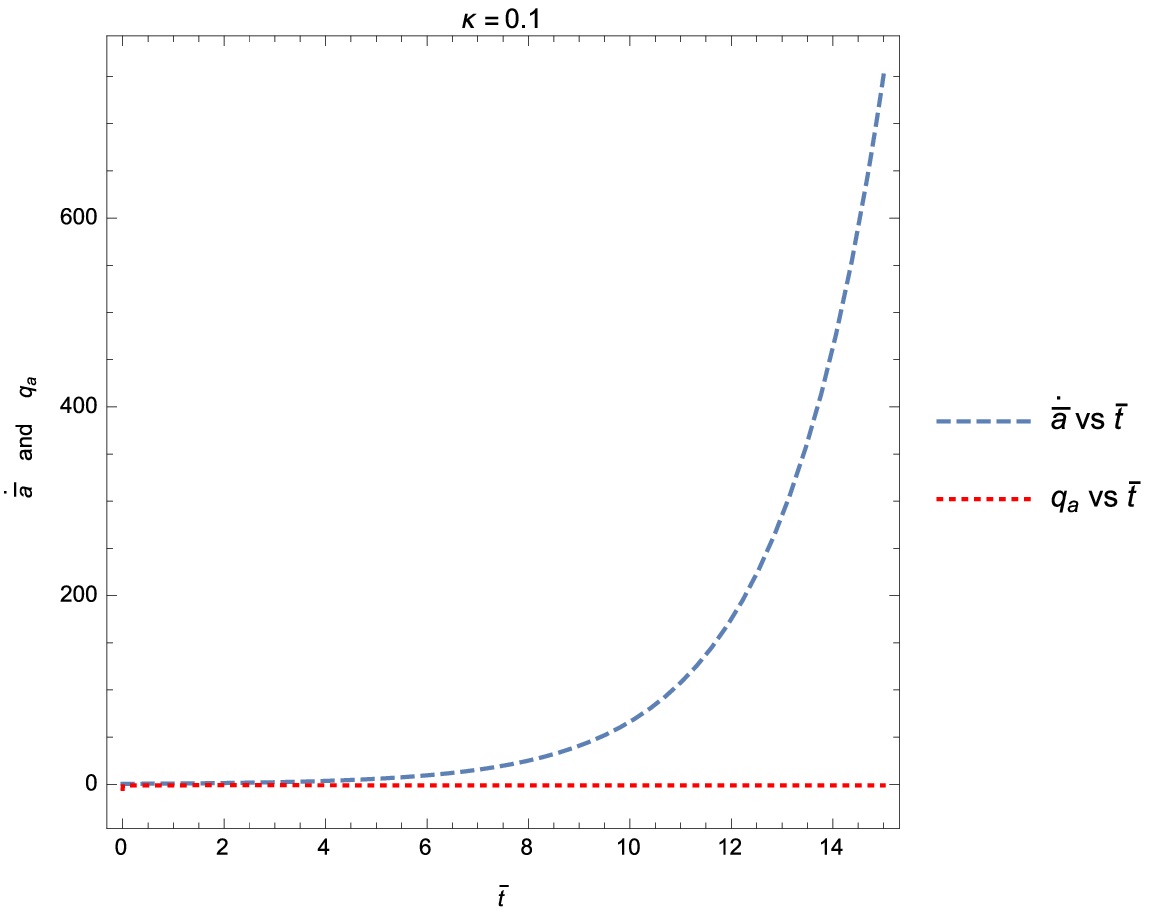}
}
\end{minipage}
\begin{minipage}[c]{0.45\textwidth}
\subfigure[\label{velacb}]{
\includegraphics[scale=0.65]{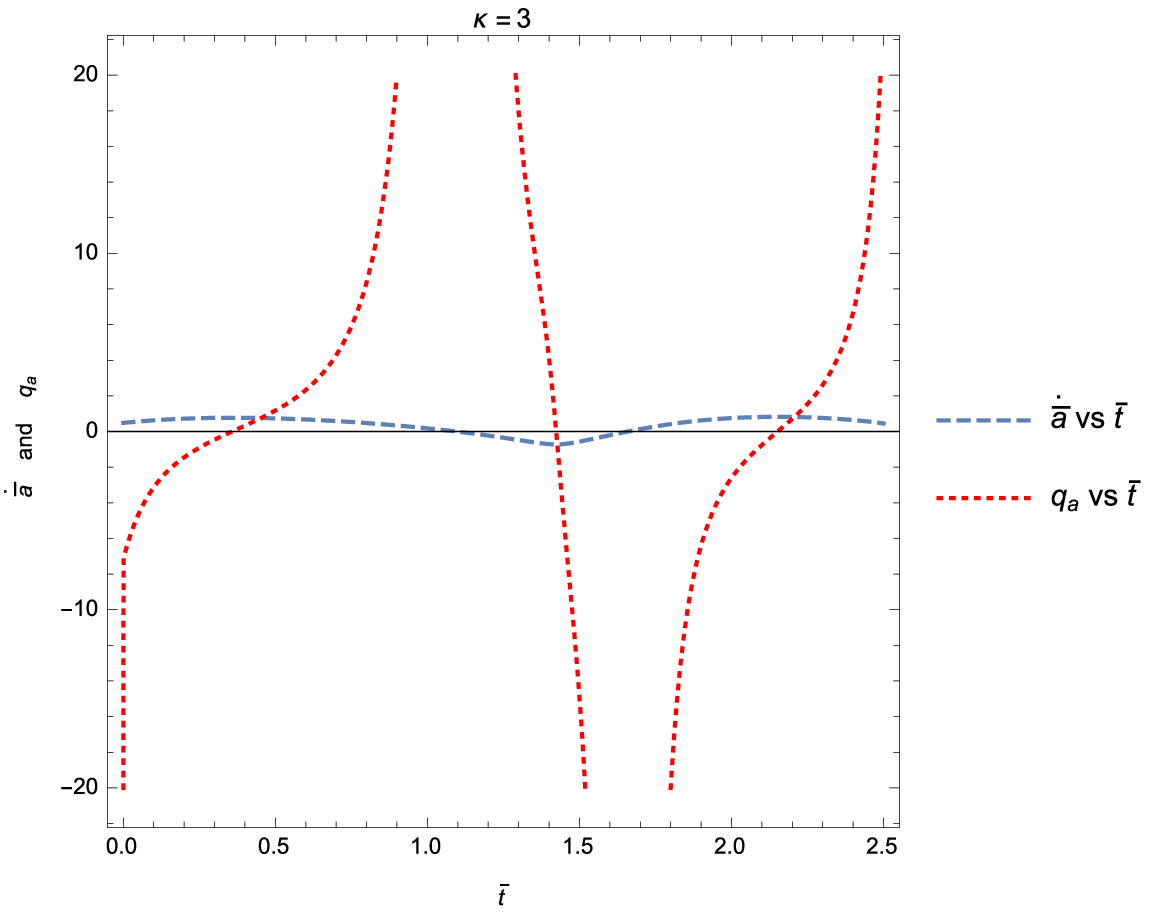}
}
\end{minipage}
\begin{minipage}[c]{0.45\textwidth}
\subfigure[\label{velacc}]{
\includegraphics[scale=0.65]{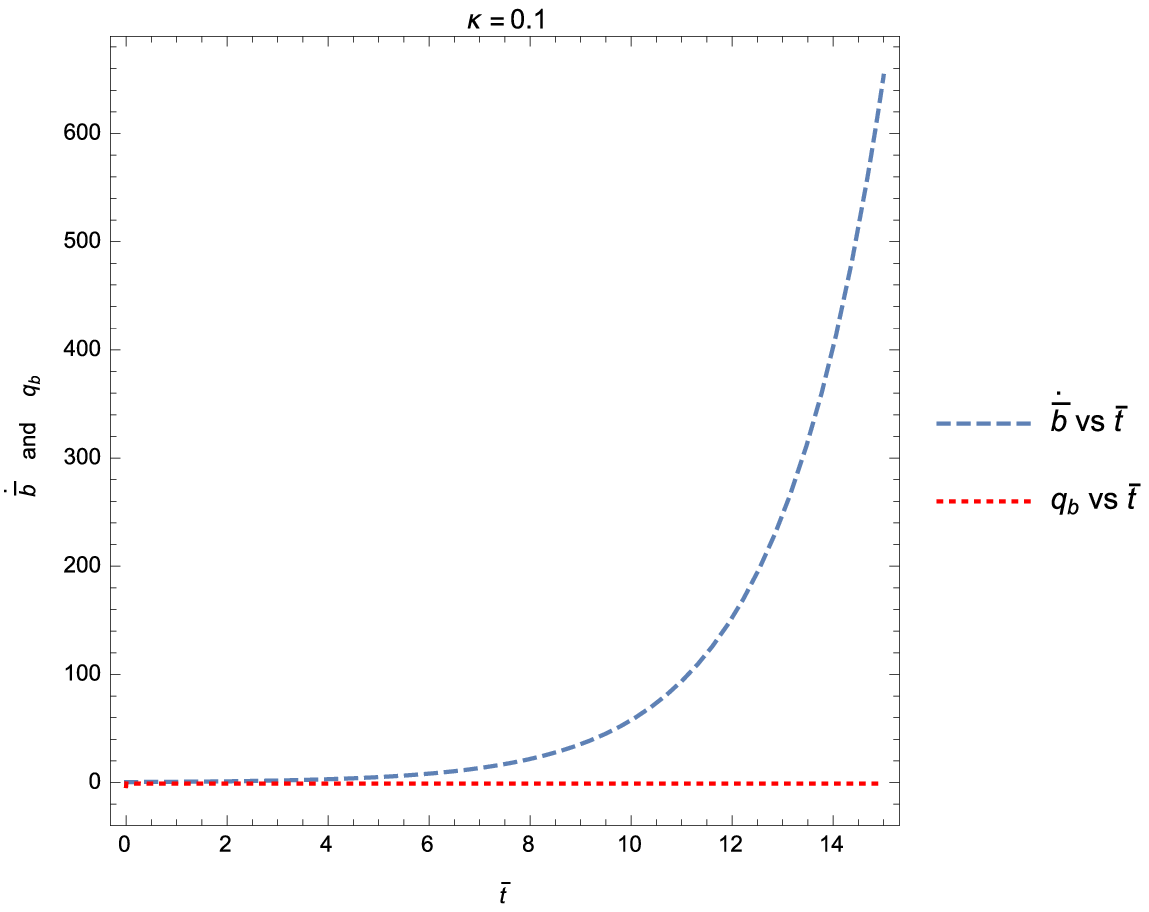}
}
\end{minipage}
\begin{minipage}[c]{0.45\textwidth}
\subfigure[\label{velacd}]{
\includegraphics[scale=0.65]{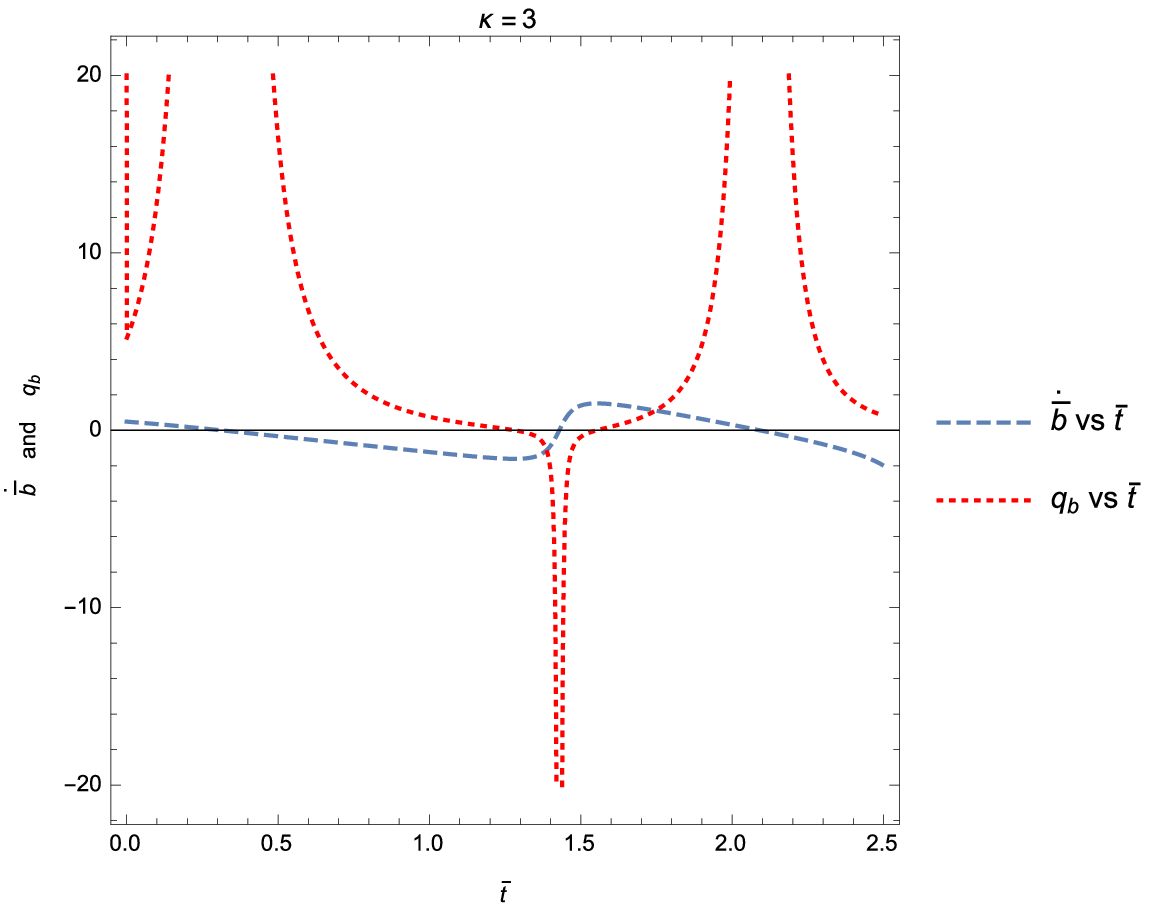}
}
\end{minipage}
\caption{\small{Expansion velocities  and deceleration parameters for
 the scale factors $\ba$ and $\bbb$ as functions of $\bt$
 in the regime  $\Lambda_a \approx \Lambda_b$ for two different values of
 $\kappa$. The panels on the right show  imprints of the quasi periodic behavior of
 scale factors.}}
\label{fig:velacsim}
\end{figure}

The  velocity  of   the  expansion  and  the  deceleration
parameter are shown in figure \ref{fig:velacsim} for the $\ba$ patch
(panels  \ref{velaca} and \ref{velacb}) and for  the $\bbb$  patch (panels
\ref{velacc} and \ref{velacd}).

For $\kappa=0.1$, the expansion velocities $\dot\ba$ and $\dot\bbb$, and the deceleration parameters $q_a$ and $q_b$, show a similar behavior,
as  expected for $\Lambda_a \approx \Lambda_b$ and initial conditions that
are symmetric under the change $\ba \leftrightarrow \bbb$.

\begin{figure}[btp]
\centering
\begin{minipage}[c]{0.45\textwidth}
\subfigure[\label{velacaneg}]{
\includegraphics[scale=0.65]{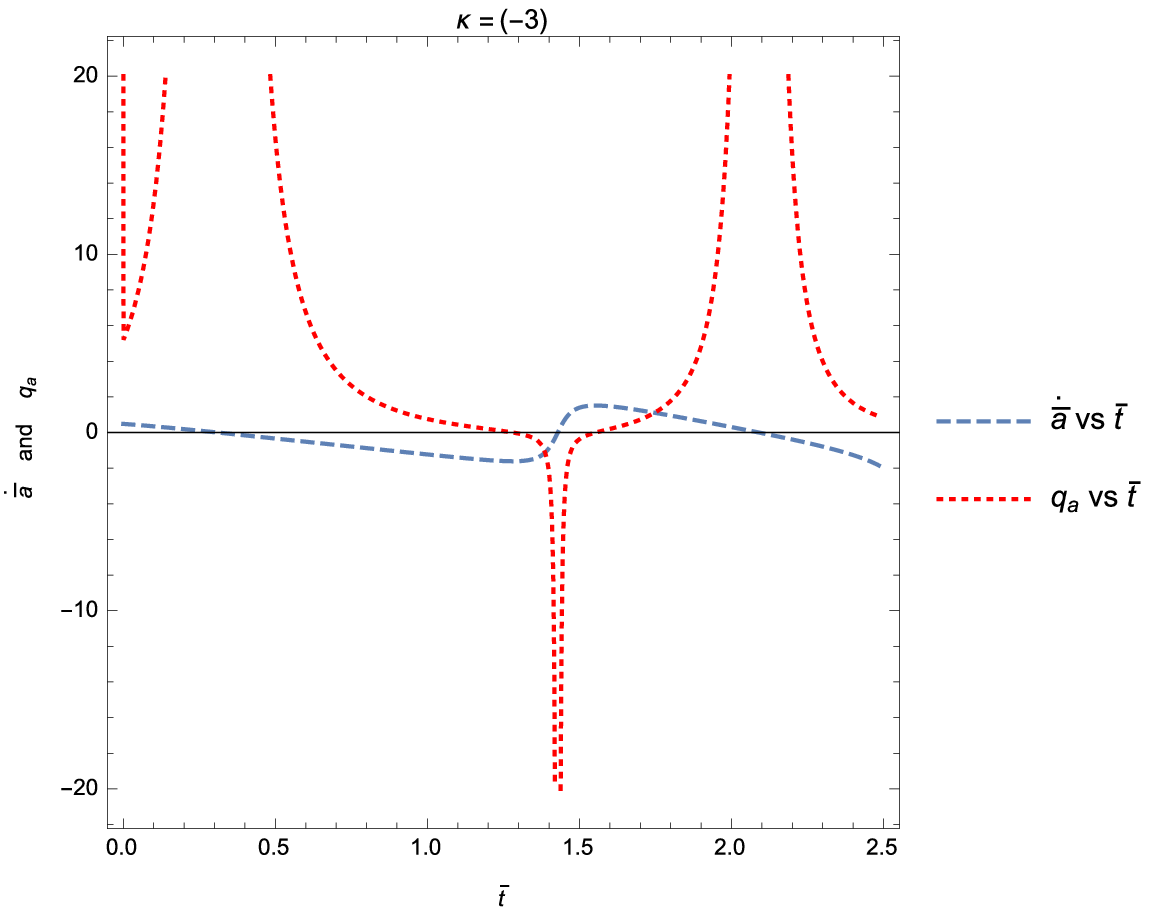}
}
\end{minipage}
\begin{minipage}[c]{0.45\textwidth}
\subfigure[\label{velacbneg}]{
\includegraphics[scale=0.65]{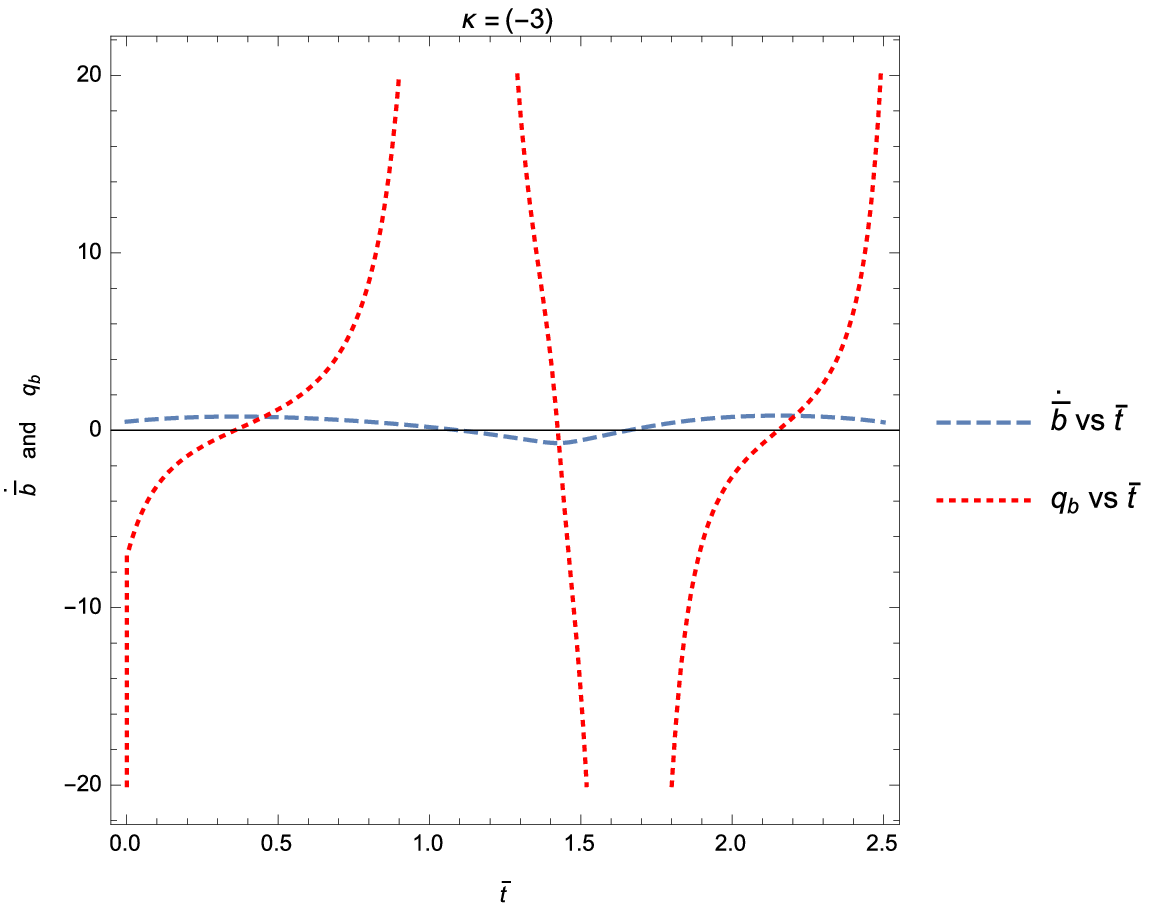}
}
\end{minipage}
\caption{\small{Expansion velocities and deceleration parameters for
 the scale factors $\ba$ and $\bbb$ as functions of $\bt$
 in the regime  $\Lambda_a \approx \Lambda_b$ for two different values of
 $\kappa<0$. We observe that panel \ref{velacb} in figure \protect\ref{fig:velacsim} goes over to panel \ref{velacbneg}
 in the present figure and panel \ref{velacd} goes over to panel \ref{velacaneg}. The asymmetry mentioned in the
 text is then explained as due to the sign of $\kappa$.}}
\label{fig:velacsimneg}
\end{figure}

For a higher value of $\kappa$ ($\kappa = 3$), the quasi periodic structure
is present although the behavior of velocity and deceleration is not symmetric
(panels \ref{velacb} and \ref{velacd}). Indeed, in the range $1.0 \alt \bt  \alt 2.3$
the scale factor $\ba$ shows two zeroes while the scale factor $\bbb$ shows three zeros (the instants for
which $q_a$ and $q_b$ go to infinity). Responsible for this asymmetry is the
fact that $\kappa$ appears with a different sign in the equations of motion for
$\ba$ and for $\bbb$.  Figure \ref{fig:velacsimneg} shows the plots for $\kappa=-3$,
clearly illustrating the change $\ba \leftrightarrow \bbb$.

%%%%%% F I G U R E 7 VER2 %%%%%%%%%%%%
%%%%%%%%%%%%%%%%%%%%%%%%%%%%%%%%

%%%%%% F I G U R E 8 VER2 %%%%%%%%%%%%
%%%%%%%%%%%%%%%%%%%%%%%%%%%%%%%%

As  a  conclusion  we  would  like  to  stress  that  the  interaction
induced through (\ref{d2}) substantially modifies the evolution of
the patches  of the universe.  For a  patch with a  small cosmological
constant compared  with the cosmological  constant of a  second patch,
interacting in the way described before, an accelerated expansion rate
is observed even  for small values of $\kappa$.  Such expansion occurs
at the expense of  the expansion of the second patch,  which  is consistent
with  the   fact  that  the   interaction  term  can  be   thought  of as an
energy-momentum source  for the  first patch,  as  can be  seen in
(\ref{conser}).

%%%%%%%%%%%%%%%%%%%%%%%%%%
\subsection{Perturbative solutions for $\kappa \ll 1$}
%%%%%%%%%%%%%%%%%%%%%%%%%%

In this  section we explore  the solutions of  the full
system of equations to first order  in a perturbative expansion in the
dimensionless parameter $\kappa$.  However, as  for the case of the Landau
problem which is formally related to  the model proposed here in view
of   the   modified  Poisson   bracket   in   (\ref{d2}),  the   limit
$\kappa \ll 1$ might be subtle.

Indeed, for the Landau problem the magnetic length $\ell = 1/\sqrt{B}$
is  ill-defined for  a magnetic  field  $B\to 0$.  Even
though  this effect  is  quantum  in nature,  our  model contains  the
dimensional parameter $\ell_g = \sqrt{G/\kappa}$, which does not admit
the limit $\kappa\to 0$. We will show  that this fact is linked to the
inflationary behavior of the  unperturbed solution and, therefore, the
perturbation theory  works well only  for times close to the initial time.  The length
$\ell_g$  is  a  clear  example where  the  non-perturbative  behavior
becomes important.

We  look  for   solutions  of  the  second  order   set  of  equations
(\ref{sec1}) and (\ref{sec2}) of the form
\eqb  \ba(\bt) &=& \ba_0
(\bt) + \kappa\, \ba_1(\bt) +{\cal O}(\kappa^2),
 \nonumber
\\
\bbb(\bt) &=& \bbb_0 (\bt) + \kappa\, \bbb_1(\bt) +{\cal O}(\kappa^2).
\label{ansatz}
\eqf
with initial condition $\ba(0)=r_a,\,\bbb(0)=r_b$.
\medskip

The zero-th  order solutions, $  \ba_0 (\bt)$ and $\bbb_0  (\bt)$, are
superpositions of functions that  contract and expand exponentially in
time.  We choose the expanding solutions
 \bb
 \ba_0     (\bt)     =     r_a\,e^{\bt\,\sqrt{\frac{\cos\lambda}{3}}},
 \,\,\,\,\,\,\,\,
\bbb_0(\bt) = r_b\, e^{\bt\, \sqrt{\frac{\sin\lambda}{3}}}.
\label{0-th}
\ee
These  solutions  satisfy  the constraint  (\ref{constdimless})  (with
$k_a=0=k_b$) and for the given initial conditions one also has
\begin{equation}
\label{pizeroth}
\dot{\ba}_0  (0) =  r_a\sqrt{\frac{\cos\lambda}{3}},\quad \dot{\bbb}_0
(0) = r_b\sqrt{\frac{\sin\lambda}{3}}.
\end{equation}

The equations at first order in $\kappa$ read
\eqb
{\ddot   \ba}_1   +   \sqrt{   \frac{\cos\lambda}{3}}  \,{\dot   \ba}_1   -
\frac{2}{3} \cos\lambda  \, \ba_1
 &=&
 \sqrt{  \frac{\sin\lambda}{3}}  \frac{r_b}{r_a} \, e^{\bt \left(   \sqrt{\frac{\sin\lambda}{3}}
    - \sqrt{\frac{\cos\lambda}{3}}\right)},
 \label{pe1}
\\
{\ddot \bbb}_1 + \sqrt{ \frac{\sin\lambda}{3}}\, {\dot \bbb}_1 - \frac{2 }{3} \sin\lambda\,
 \bbb_1 &=& -\sqrt{ \frac{\cos\lambda}{3}} \,   \frac{r_a}{r_b} \,
 e^{-\bt\left(\sqrt{\frac{\sin\lambda}{3}}
    -\sqrt{\frac{\cos\lambda}{3}}\right)}.
\label{pe2}
\\
r_a^2\,\sqrt{\cos\lambda}\, e^{2 \bt  \sqrt{\frac{\cos\lambda}{3}}}\left[
\dot{\ba}_1 -\sqrt{\frac{\cos\lambda}{3}}\, \ba_1\right]
& = &  -
r_b^2\,\sqrt{\sin\lambda}\, e^{2 \bt  \sqrt{\frac{\sin\lambda}{3}}}\left[
\dot{\bbb}_1 - \sqrt{\frac{\sin\lambda}{3}}\, \bbb_1\right]
\label{pe3}
\eqf

The solutions of the set of equations (\ref{pe1}) and (\ref{pe2})
depend   on    two   arbitrary    constants   when    the   conditions
$\ba_1(0) =  0 =  \bbb_1(0)$ are imposed.  Initial conditions  for the
velocities are  chosen in agreement with  (\ref{f3}), which guarantees
the     constraint     and     then    $\dot{\ba}(0)     =     0     =
\dot{\bbb}(0)$. The integration  constants are now fixed  and the solutions to
order $\kappa$ are
\begin{eqnarray}
\label{asmallkappa}
\ba(\bt) &=& e^{\bt\sqrt{\frac{\cos\lambda}{3}}} \,r_a +
e^{-2\bt\,\sqrt{\frac{\cos\lambda}{3}}}\left(\frac{\kappa\,r_b}{r_a}\right)\sqrt{\frac{\sin\lambda}{3}}
\left(\frac{\sec\lambda}{2+\sqrt{\tan\lambda}-\tan\lambda}\right)
\bigg[2+e^{\bt\sqrt{3\cos\lambda}}-3e^{\bt\sqrt{\frac{\cos\lambda}{3}}\left(1+\sqrt{\tan\lambda}\right)}+
 \nonumber
 \\
 &&
\sqrt{\tan\lambda}\left(e^{\bt\sqrt{3\cos\lambda}}-1\right)\bigg] ,
 \\
 \label{bsmallkappa}
 \bbb(\bt)&=& e^{\bt\sqrt{\frac{\sin\lambda}{3}}} \,r_b -
 e^{-2\bt\,\sqrt{\frac{\sin\lambda}{3}}}\left(\frac{\kappa\,r_a}{r_b}\right)\frac{1}{\sqrt{3\sin\lambda}}
\left(\frac{1}{2\tan\lambda+\sqrt{\tan\lambda}-1}\right)
\bigg[ e^{\bt\sqrt{3\cos\lambda}}-1 +
\nonumber
\\
&&\left( 2+e^{\bt\sqrt{3\sin\lambda}}-3e^{\bt\sqrt{\frac{\cos\lambda}{3}}\left(1+\sqrt{\tan\lambda}\right)}
\right) \sqrt{\tan\lambda}\bigg] .
\end{eqnarray}

These  are  the scale  factors  of  the two  patches  when  a kind  of
interaction is introduced by a  modification of the Poisson brackets
of momenta in  the $\kappa \ll 1$ limit. These  solutions satisfy the
initial                                                     conditions
$\ba(0)=r_a, \bbb(0)=r_b, \dot{\ba}(0)\neq0, \dot{\bbb}(0)\neq0$.

\begin{figure}[H]
\begin{center}
\includegraphics[width=.5\textwidth]{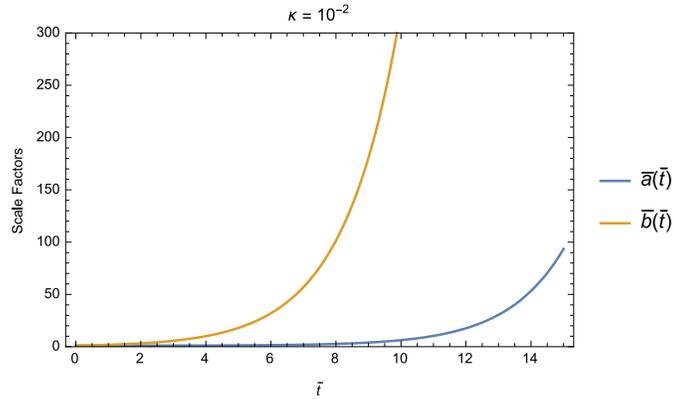}\\
\caption{Scale factors $\ba(\bt)$ and $\bbb(\bt)$ for $\kappa=10^{-2}$
  and  $\lambda=\pi/2-10^{-4}$.    One can  observe  the
  exponential growth of $\ba$.}
  \label{fig:fig7}
   \end{center}
\end{figure}

Let us explore the case $\Lambda_a\sim 0$ (or $\lambda\sim\pi/2$).  In
figure   \ref{fig:fig7}    we   observe   the   scale    factors   for
$\kappa =10^{-2}$ and $\lambda =\pi/2 -10^{-4}$.
%%%%%%%%%%%%%%%%%%%%%%
%%%%% F I G U R A 7 %%%%%%%%%%
%%%%%%%%%%%%%%%%%%%%%%

\begin{figure}[H]
\centering
\begin{minipage}[c]{0.45\textwidth}
\subfigure{
\includegraphics[scale=0.7]{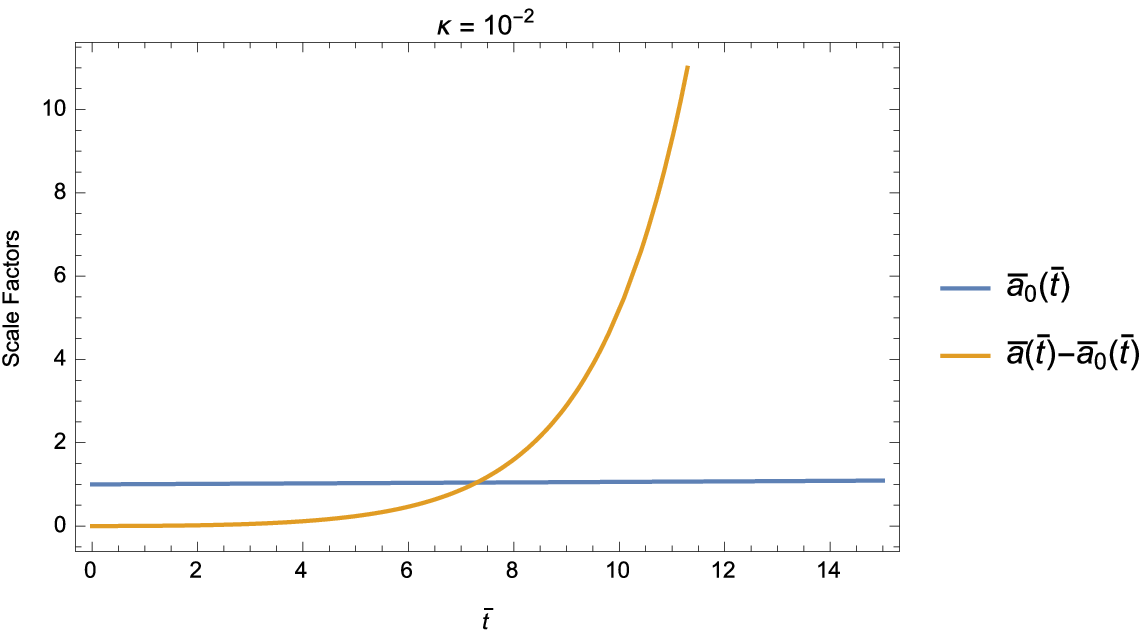}
\label{fig-adela}
}
\end{minipage}
\begin{minipage}[c]{0.5\textwidth}
\subfigure{
\includegraphics[scale=0.7]{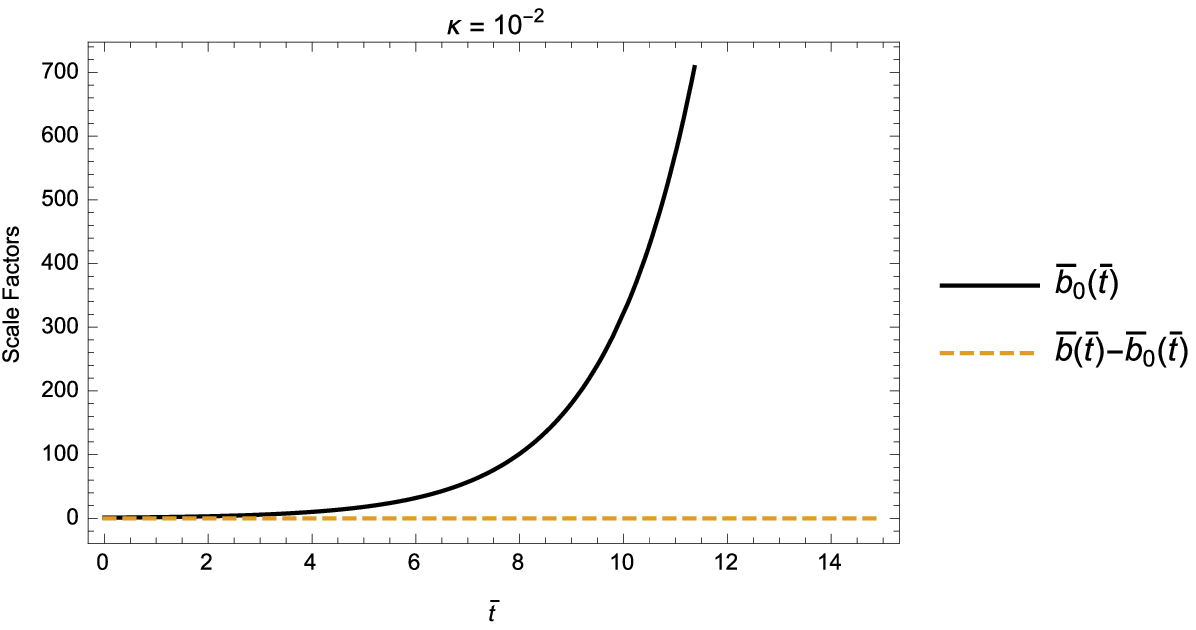}
\label{fig-adelb}
}
\end{minipage}
\caption{\small{The  perturbative  terms  grow exponentially  for  the
    scale factor $\ba$  for $\kappa \ll 1$. The scale factor  $\bbb$ does not
    show the same  behavior.  The perturbative  approach is  valid, therefore,
    only at  early stages in the evolution of the system.}}
\label{fig:fig8}
\end{figure}

It  can be  seen in  (\ref{asmallkappa}) and  (\ref{bsmallkappa}) that
perturbative  terms  contain an  exponential   dependence  in  time  and
therefore  the   perturbative  approach   is  valid  only   for  early
times.  Indeed, in  Figure  \ref{fig:fig8} we  can  compare the  scale
factors with  $\kappa=0$, with the  behavior of the  perturbative term
$\Delta\ba \equiv  \ba(\bt) -  \ba_0(\bt)$ (and the  analoguous definition
for  $\Delta\bbb$).  Panel  \ref{fig-adela} shows  the case  of $\ba$,
while panel \ref{fig-adelb} shows the case of $\bbb$.

%%%%%%%%%%%%%%%%%%%%%%
%%%%% F I G U R A 8 %%%%%%%%%%
%%%%%%%%%%%%%%%%%%%%%%
%%%%%%%%%%%%%%%%%%%%%%%%

The perturbative approach in the  present case (see figure \ref{fig:fig8})
is  valid  for times  $\bt  <8$.  But  we  observe that,  already  for
$\bt\approx 4$, the scale factor  $\ba$ starts growing. In this sense,
we say  that the perturbation  solution is  valid at early  times. The
precise  value of  the time  cutoff depends  on $\kappa$,  as well  as
$\lambda$.

\begin{figure}[H]
\begin{center}
\includegraphics[width=.5\textwidth]{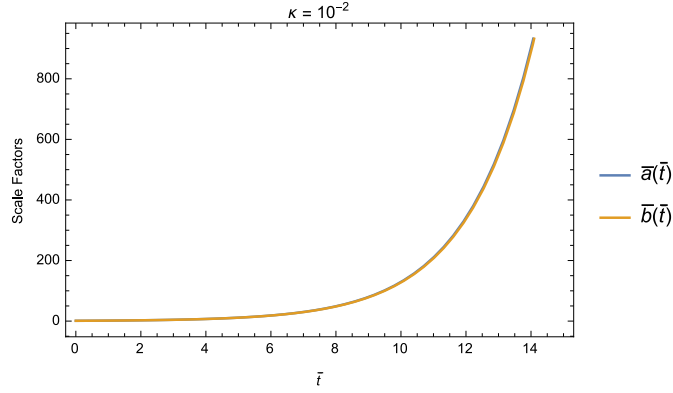}\\
\caption{Comparison of scale factors $\ba(\bt)$ and $\bbb(\bt)$ for $\kappa=10^{-2}$
  and $\lambda=\pi/4-10^{-4}$.}
  \label{fig:fig9}
   \end{center}
\end{figure}

The second case of interest is $\Lambda_a\sim\Lambda_b$. Scale factors
are shown  in figure \ref{fig:fig9}  and one can observe that they are almost equal, as expected.

%%%%%%%%%%%%%%%%%%%%%%
%%%%% F I G U R A 9 %%%%%%%%%%
%%%%%%%%%%%%%%%%%%%%%%

\begin{figure}[H]
\centering
\begin{minipage}[c]{0.45\textwidth}
\subfigure{
\includegraphics[scale=0.7]{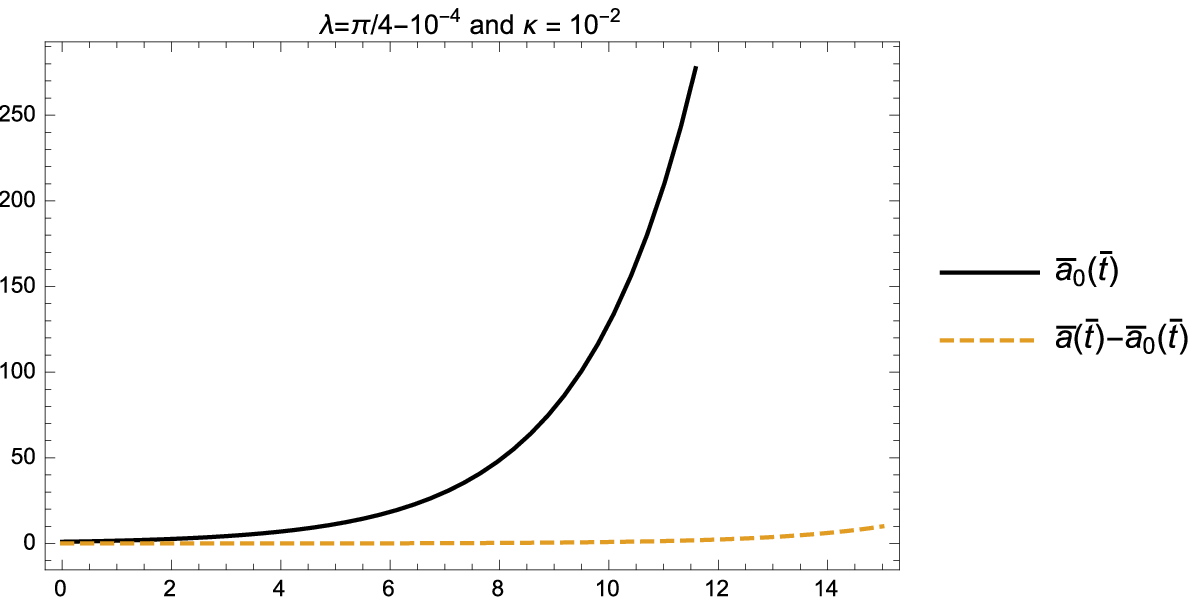}
\label{fig-adelasim}
}
\end{minipage}
\begin{minipage}[c]{0.4\textwidth}
\subfigure{
\includegraphics[scale=0.7]{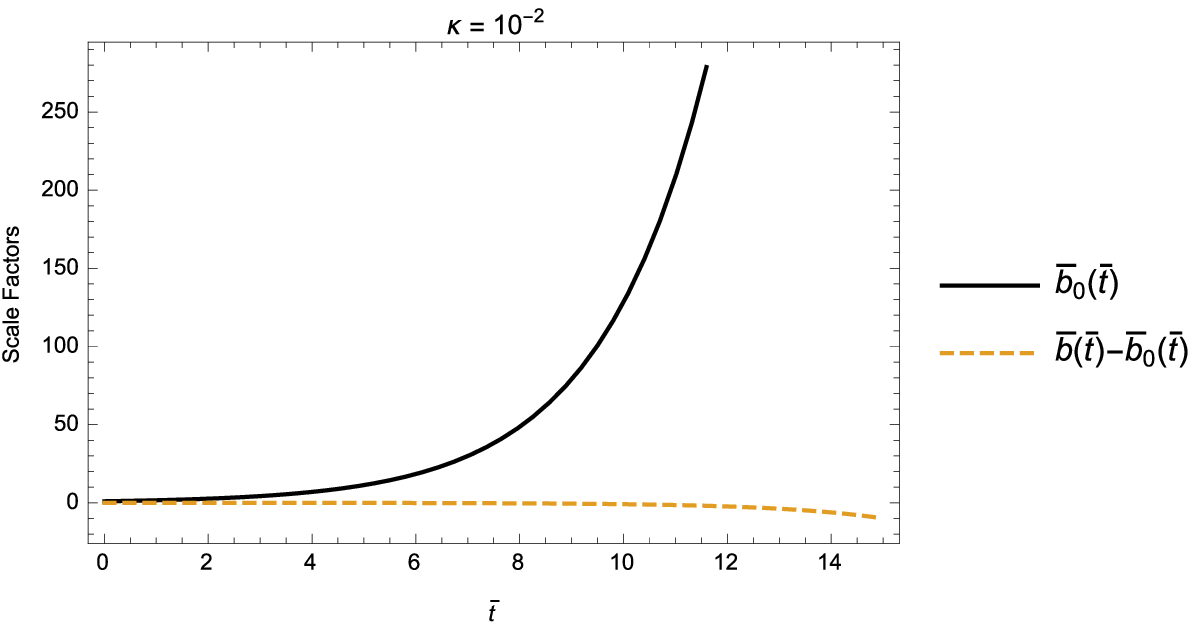}
\label{fig-adelbsim}
}
\end{minipage}
\caption{\small{For  $\Lambda_a  \sim  \Lambda_b$,  perturbations  are
    always  smaller  than  the  unperturbed  solution.  Here  we  show
    $\lambda =  \pi/4 -10^{-4}$ and $\kappa=10^{-2}$.   }}
\label{fig:fig10}
\end{figure}

To check  the validity  of the perturbation  expansion, we  plot the scale
factors with  $\kappa=0$ (unperturbed solutions) and  compare them with the
perturbed solutions.  Figure \ref{fig:fig10}  shows the  results. We  observe
that for the present case, no early time restrictions are present and
therefore  the   perturbation  expansion  can  be   safely  applied  for
$\kappa \ll 1$.

%%%%%%%%%%%%%%%%%%%%%%
%%%%% F I G U R A 10 %%%%%%%%%%
%%%%%%%%%%%%%%%%%%%%%%
%%%%%%%%%%%%%%%%%%%%%%%%

In  any case,  we  observe  that the  $\bbb$  scale  factor induces  a
cosmological constant  on the universe  described by the  $\ba$ scale factor, {\it i.e.},  a sort  of dark  energy \cite{frieman}  coming from
another sector (or patch) of the Universe.

The idea that causally disconnected regions influence the evolution of
some  part of  the universe  by assuming  a small  interaction between the
regions is  an  interesting proposal in itself.  The solutions discussed
in this paper show that the universe would experience inflation faster than in  the cosmological  standard model
(or  slower,  depending on the  initial
conditions). More
importantly,   even  if  the cosmological constant  $\Lambda_a$ of our patch of the Universe were  very  small at  the
present epoch, our patch may actually  accelerate
\cite{perl,riess} by the  presence of a second  patch with nonzero cosmological constant $\Lambda_b$
under the Poisson-bracket interaction proposed here.

%%%%%%%%%%%%%%%%%%%%
%%%%%%%%%%%%%%%%%%%%
\section{Dark Energy}
%%%%%%%%%%%%%%%%%%%%
%%%%%%%%%%%%%%%%%%%%

As discussed in Section II, the coupling between two causally disconnected
regions induces interactions that are
 consistent with energy-momentum conservation. An observer in the $a$ patch would measure an effective density $\rho^{(a)}$ and an effective pressure $p^{(a)}$ as given  in Eqs.\ \eqref{conser}.
 
Knowing  the  solutions for $a(t)$  and  $b(t)$, this model  can
predict whether there is  dark energy effect or not. We can numerically compute   the
pressure $p^{(a)}$, the density $\rho^{(a)}$  and  evaluate from the  equation of state the parameter $w$.

In particular, it follows that
\bb
p^{(a)}+\rho^{(a)} = \frac{1}{8\pi G} \left[ \Lambda_b \left(\frac{b}{a}\right)^3 -
3 \frac{b}{a^3}\left( {\dot b}^2   + k_b\right)-2 \kappa\,\frac{{\dot b}}{a^2} \right]. \label{state2}
\ee
For dark energy, one has empirically that $p+\rho \approx 0$. Therefore in the context we have been discussing in the present work,
 if the solutions of the Einstein equations 
 become quasi periodic, they acquire a behavior compatible with a dark energy component
 only during some specific periods of the evolution.

In other words, the quasi oscillatory nature of the cosmological solutions we have found eventually stops the accelerated expansion of the Universe and, from this point of view, the concept of dark energy as a particular component of the Universe could become unnecessary.

%%%%%%%%%%%%%%%%%%%%
%%%%%%%%%%%%%%%%%%%%
\section{Cocycles}
%%%%%%%%%%%%%%%%%%%%
%%%%%%%%%%%%%%%%%%%%

The mathematics of Poisson bracket deformation is well known~\cite{flato,catta,wilde,fedosov,flato1,kosen} and as was  previously mentioned, the deformed Poisson bracket (\ref {d2}) is reminiscent of the magnetic translations group in the quantum Hall effect. A magnetic translation is an operator ${\hat T}[{\theta}]$ defined by \cite{zak}
\bb
{\hat T}[{\theta}] = e^{i {\bf \pi}_a\, {\theta^a}}, \label{tras1}
\ee
where $\pi_a$ are the momenta defined in Section I and ${\theta^a}$ are real parameters.
In terms of this notation, the internal composition law of the magnetic group is very unconventional because it satisfies
\bb
{\hat T}[{\theta }]\, {\hat T}[{\tau}] = e^{\frac{i\kappa}{2} \epsilon_{ab}\theta^a\tau^b} {\hat T} (\theta
+\tau). \label{tras2}
\ee

The phase in (\ref{tras2}) is a 3-cocycle implying that the generators ${\hat T}[{\theta}]$  are a ray representation of the group of {\bf magnetic translations} \cite{freidel3}. A good example of this are the Gauss's anomalies terms of cocycles found by Faddeev in {\bf 1984} \cite{faddeev,jackiw}.

From a conceptual point of view the deformation of the Poisson bracket includes not only a crucially important change (namely, the appearance of quasi-periodicity) 
but also makes explicit the presence of 3-cocycles which were not obvious a priori.

%%%%%%%%%%%%%%%%%%%%%
%%%%%%%%%%%%%%%%%%%%%
\section{Summary and Discussion}
%%%%%%%%%%%%%%%%%%%%%
%%%%%%%%%%%%%%%%%%%%%
In this paper we have studied an extension of the FLRW model with two metrics assuming that two  patches of the universe
(causally disconnected in principle) interact  through  a modification of the Poisson bracket of the momenta of the two
 metrics, one on each patch. This modification is inspired by an analogy with the quantum Hall effect. Following
 this analogy, the deformation parameter of the Poisson bracket might be interpreted as a  minimum distance between two neighboring regions that rotate  with respect to a plane of the target two-metric space  or, in other words, this assumption is equivalent to the analog of the lowest Landau level in cosmology.

However, the assumption explained above heuristically is not exempt from mathematical subtleties, the first as explained above, assumes that the deformation (\ref{d2}) corresponds to a change of algebraic structure known as group of magnetic translations which contains a internal composition rule containing a 3-cocycle.
This 3-cocycle is responsible for the oscillatory behavior of the equations of
FLRW we have found in this paper.

Finally we would like to insist on the fact that  the quasi periodic  structure of the solutions of the extended FLRW equations contain both acceleration and deceleration epochs, and therefore if these quasi periodic solutions are used for interpreting dark energy, it would mean that the observations of the current universe \cite{perl} and \cite{riess} are only a snapshot of the universe evolution at a particular (accelerating) time.

However, there is a highly non-trivial phenomenon in the results presented here. In the theory discussed in
this paper there are inflationary solutions with negligible cosmological constant in our patch of the universe ($\Lambda_a\ll\Lambda_b$). This fact is a consequence of the causality breaking produced by the deformation of the
momenta of the metric.

\begin{acknowledgments}
One of us (J.G.) thanks to J. L. Cortes for  discussions. This  work  was supported  by  FONDECYT/Chile  grants 1130020  (J.G.),
1140243 (F.M.) and USA-155 (J.G.). P. G was partially supported by NSF award PHY-1415974 at the
University of Utah. H.F. thanks ANPCyT, CONICET and UNLP, Argentina, for partial support through grants PICT-2014-2304, PIP 2015-2017 GI -688CO and Proy. Nro. 11/X748, respectively.
\end{acknowledgments}

%%%%%%%%%%%%%%%%%%%%%%%%%%%%%%%%%%%%%%%%%%%%%%%%%%%%%%%%%%%%
%%%%%%%%%%%%%%%%%%%%%%%%%%%%%%%%%%%%%%%%%%%%%%%%%%%%%%%%%%%%
%%%%%%%%%%%%%%%%%%%%%%%%%%%%%%%%%%%%%%%%%%%%%%%%%%%%%%%%%%%%

{\appendix

\section {Dimensions}

We consider the Hilbert-Einstein action with cosmological constant
\begin{equation}
\label{h-e}
S = \frac{1}{16\pi G}\int \sqrt{-g}(R+\Lambda)d^4x .
\end{equation}
In natural  units the canonical  dimensions of the  different elements
composing the  action are  $[G] =-2$, $[g]=0$,  $[R]=+2$, $[d^4x]=-4$,
$[\Lambda] = 2$, so that the action has canonical dimension zero.

For the FLRW metric, the scale factor $a$ and the lapse function $N$ can be chosen as dimensionless.
 Since the scale factor depends on time only, the
action (\ref{h-e}) reduces to
\begin{equation}
\label{aa}
S = \frac{V}{16\pi G}\int \sqrt{-g(t)}(R(t)+\Lambda)dt
\end{equation}
where $V$ is a space volume. The quantity $V/G$ has canonical  dimension $-1$. We
define a quantity ${ v}=V/(16\pi G)$ and write the action as
\begin{equation}
\label{aa-part}
S = {v}\int{\mathcal L}(a,\dot{a})dt.
\end{equation}
The canonical dimension of $ v{\cal L} $  is $+1$. In order to have a close
analogy  with  the  case  of the Landau problem  (which  is  defined  for
particles rather than fields) we consider the Lagrangian
$L\equiv v {\cal L}$ and the action
$$
S=\int L\,dt
$$
Since $a$  is dimensionless  and $L$ has  canonical dimension  $+1$, the
canonical  momentum   $p_a=\partial  L/\partial  \dot  {a}$   is  also
dimensionless. The Poisson bracket between coordinates and momenta -- inherited from the gravity theory as
a field theory -- is also dimensionless as it should be
$$
\{a,p_a\}=1.
$$
For  the case  of particles,  the  coordinates and  canonical
momenta have inverse dimensions to each other, and while the Poisson bracket between coordinates and momenta is
dimensionless, the  modified  Poisson   bracket  between  momenta  has
canonical dimension $+2$.

In our case one  can do a similar choice. Indeed, let  us define a new
variable $\tilde{a} =a \sqrt{G}$ with canonical dimensions $-1$ (as the spatial coordinates in a particle theory),
then the canonical momentum will have dimensions $+1$.

Since ${\cal L}$ is a homogeneous function of $a$ of degree 3
$$
L=v{\cal L}(a,\dot{a})=\frac{v}{G^{3/2}}{\cal L} (\tilde{a},\dot{\tilde{a}})=\frac{V}{16\pi G^{3/2}}\frac{1}{G}
{\cal L}(\tilde{a},\dot{\tilde{a}})
$$
Finally, the quantity $V/(16\pi G^{3/2})$ is a dimensionless constant, and therefore we can take as the  Lagrangian
for our model
\begin{equation}
L\equiv  \frac{1}{G}{\cal L}(\tilde{a},\dot{\tilde{a}}).
\end{equation}
The action is thus redefined up to a dimensionless parameter.  The Lagrangian so redefined has canonical dimension +1,
$\tilde{a}$ has dimension $-1$ and ${\tilde{p}_a}$ has dimension $+1$. The deformed Poisson bracket then
has dimension $+2$, as desired.

In the text we use $a$ instead of $\tilde{a}$ for simplicity.
}


\begin{thebibliography}{99}
%%%%%%%%%%%%%%%%%%%%%%%%%%%%%%%%%%%%%%%%%%%%%%%%%%%%%%%%%%%%
\bibitem{Linde}  For a beautiful historic description about this,  see  A.~Linde, ``A brief history of the multiverse,''
  arXiv:1512.01203 [hep-th].
  %%%%%%%%%%%%%%%%%%%%%%%%%%%%%%%%%%%%%%%%%%%%%%%%%%%%%%%%%%%%
\bibitem{Kibble}
  T.~W.~B.~Kibble,
  %``Topology of Cosmic Domains and Strings,''
  J.\ Phys.\ A {\bf 9}, 1387 (1976).

%%%%%%%%%%%%%%%%%%%%%%%%%%%%%%%%%%%%%%%%%%%%%%%%%%%%%%%%%%%%
%%
%%%%%%%%%%%%%%%%%%%%%%%%%%%%%%%%%%%%%%%%%%%%%%%%%%%%%%%%%%%%
%%%%%%%%%%%%%%%%%%%%%%%%%%%%%%%%%%%%%%%%%%%%%%%%%%%%%%%%%%%%
%%%%%%%%%%%%%%%%%%%%%%%%%%%%%%%%%%%%%%%%%%%%%%%%%%%%%%%%%%%%
  \bibitem{Vilenkin}
  A.~Vilenkin,
  %``Cosmic Strings and Domain Walls,''
  Phys.\ Rept.\  {\bf 121}, 263 (1985).

%%%%%%%%%%%%%%%%%%%%%%%%%%%%%%%%%%%%%%%%%%%%%%%%%%%%%%%%%%%%
%%%%%%%%%%%%%%%%%%%%%%%%%%%%%%%%%%%%%%%%%%%%%%%%%%%%%%%%%%%%
  \bibitem{monopole}
  B.~Kleihaus, J.~Kunz and Y.~Shnir,
  %``Monopole-antimonopole chains and vortex rings,''
  Phys.\ Rev.\ D {\bf 70} (2004) 065010
  %%%%%%%%%%%%%%%%%%
  \bibitem{Starobinsky} A.~A.~Starobinsky,
  %``A New Type of Isotropic Cosmological Models Without Singularity,''
  Phys.\ Lett.\  {\bf 91B}, 99 (1980); ibid. ,
  %``Spectrum of relict gravitational radiation and the early state of the universe,''
  JETP Lett.\  {\bf 30}, 682 (1979)
  [Pisma Zh.\ Eksp.\ Teor.\ Fiz.\  {\bf 30}, 719 (1979)].%%%%%%%%%%%%%%%%%%%%%%%%%%%%%%%%%%%%%%%%%%%%%%%%%%%%%%%%%%%%
  %%%%%%%%%%%%%%%%%%%%%
  \bibitem{guth}
  A.~H.~Guth,
  %``The Inflationary Universe: A Possible Solution to the Horizon and
  %Flatness Problems,''
  Phys.\ Rev.\ D {\bf 23}, 347 (1981).

  \bibitem{linde}
  A.~D.~Linde,
  %``A New Inflationary Universe Scenario: A Possible Solution of the
  %Horizon, Flatness, Homogeneity, Isotropy and Primordial Monopole Problems,''
  Phys.\ Lett.\ B {\bf 108}, 389 (1982); ibid,  Phys.\ Lett.\ B {\bf 114}, 431 (1982),
  %``Chaotic Inflation,''
  Phys.\ Lett.\ B {\bf 129}, 177 (1983).
  %%%%%%%%%%%%%%%%%%%%%%%%%%%%%%%%%%%%%%%%%%%%%%%%%%%%%%%%%%%%
%%%%%%%%%%%%%%%%%%%%%%%%%%%%%%%%%%%%%%%%%%%%%%%%%%%%%%%%%%%%
  \bibitem{steinhardt} A.~Albrecht and P.~J.~Steinhardt,
  %``Cosmology for Grand Unified Theories with Radiatively Induced Symmetry Breaking,''
  Phys.\ Rev.\ Lett.\  {\bf 48}, 1220 (1982).

%%%%%%%%%%%%%%%%%%%%%%%%%%%%%%%%%%%%%%%%%%%%%%%%%%%%%%%%%%%%
\bibitem{linde1} A.~D.~Linde,
  %``The Inflationary Universe,''
  Rept.\ Prog.\ Phys.\  {\bf 47} (1984) 925.
\bibitem{linde2} A.~D.~Linde,
  %``Particle physics and inflationary cosmology,''
  Contemp.\ Concepts Phys.\  {\bf 5} (1990) 1
\bibitem{linde3}A.~D.~Linde,
  %``Inflationary Cosmology,''
  Lect.\ Notes Phys.\  {\bf 738} (2008) 1.

 \bibitem{mukhanov} V.~F.~Mukhanov,
  %``Gravitational Instability of the Universe Filled with a Scalar Field,''
  JETP Lett.\  {\bf 41}, 493 (1985)
  [Pisma Zh.\ Eksp.\ Teor.\ Fiz.\  {\bf 41}, 402 (1985)].
  %%%%%%%%%%%%%%%%%%
  \bibitem{kazana} D.~Kazanas, Astrophys. J. \ {\bf 241}, L59 (1980).
  %%%%%%%%%%%%%%%%%%%%%%%%%%%%%%%
   \bibitem{sato} K.~Sato,
  %``First Order Phase Transition of a Vacuum and Expansion of the Universe,''
  Mon.\ Not.\ Roy.\ Astron.\ Soc.\  {\bf 195}, 467 (1981).

  \bibitem{Type-Ia} A.~G.~Riess {\it et al.} [Supernova Search Team],
  %``Type Ia supernova discoveries at z > 1 from the Hubble Space Telescope: Evidence for past deceleration and constraints on dark energy evolution,''
  Astrophys.\ J.\  {\bf 607}, 665 (2004).

  %%%%%%%%%%%%%%%%%%%%%%%%%%%%%%%%%%%%%%%%%%%%%%%%%%
  \bibitem{WMAP} D.~N.~Spergel {\it et al.} [WMAP Collaboration],
  %``First year Wilkinson Microwave Anisotropy Probe (WMAP) observations: Determination of cosmological parameters,''
  Astrophys.\ J.\ Suppl.\  {\bf 148}, 175 (2003).
    %%%%%%%%%%%%%%%%%%%%%%%%%%%%%%%%%%%%%%%%%%%%%%%%%%
  \bibitem{Weinberg}S.~Weinberg,
  %``The Cosmological Constant Problem,''
  Rev.\ Mod.\ Phys.\ {\bf 61} (1989) 1.
  \bibitem{Zeldovich}  Y.~B.~Zeldovich, and A.~Krasinski ,
  %``The Cosmological constant and the theory of elementary particles,''
  Sov.\ Phys.\ Usp.\  {\bf 11}, 381 (1968)
  [Gen.\ Rel.\ Grav.\  {\bf 40}, 1557 (2008)]
  [Usp.\ Fiz.\ Nauk {\bf 95}, 209 (1968)]; Y.~B.~Zeldovich,
  %``Cosmological Constant and Elementary Particles,''
  JETP Lett.\  {\bf 6}, 316 (1967)
  [Pisma Zh.\ Eksp.\ Teor.\ Fiz.\  {\bf 6}, 883 (1967)]; S.~M.~Carroll,
  %``The Cosmological constant,''
  Living Rev.\ Rel.\  {\bf 4}, 1 (2001).
  %%%%%%%%%%%%%%%%%%%%%%%%%%%%%%%%%%%%%%%%%%%%%%%%%%%%%%%%%%%%
%%%%%%%%%%%%%%%%%%%%%%%%%%%%%%%%%%%%%%%%%%%%%%%%%%%%%%%%%%%%
  \bibitem{Carroll}
  S.~M.~Carroll,
  %``The Cosmological constant,''
  Living Rev.\ Rel.\  {\bf 4}, 1 (2001)
  %%%%%%%%%%%%%%%%%%%%%%%%%%%%%%%%%%%%%%%%%%%%%%%%%%%%%%%%%%%
  %%%%%%%%%%%%%%%%%%%%%%%%%%%%%%%%%%%%%%%%%%%%%%%%%%%%%%%%%%%%%
 \bibitem{peebles}  P.~J.~E.~Peebles and B.~Ratra,
  %``The Cosmological constant and dark energy,''
  Rev.\ Mod.\ Phys.\  {\bf 75} (2003) 559

%%%%%%%%%%%%%%%%%%%%%%%%%%%%%%%%%%%%%%%%%%%%%%%%%%%%%%%%%%%%
\bibitem{wetterich} C.~Wetterich,
  %``Cosmology and the Fate of Dilatation Symmetry,''
  Nucl.\ Phys.\ B {\bf 302}, 668 (1988).
  %%%%%%%%%%%%%%%%%%%%%%%%%%%%%%%%%%%%%%%%%%%%%%%%%%%%%%%%%%%
  \bibitem{dvali} See for example, C.~Deffayet, G.~R.~Dvali and G.~Gabadadze,
  %``Accelerated universe from gravity leaking to extra dimensions,''
  Phys.\ Rev.\ D {\bf 65}, 044023 (2002)
  %%%%%%%%%%%%%%%%%%%%%%%%%%%%%%%%%%%%%%%%%%%%%%%%%%%%%%%%%%
%%%%%%%%%%%%%%%%%%%%%%%%%%%%%%%%%%%%%%%%%%%%%%%%%%%%%%%%%%%%

\bibitem{Chang:2001bm}
  L.~N.~Chang, D.~Minic, N.~Okamura and T.~Takeuchi,
  %``The Effect of the minimal length uncertainty relation on the density of states and the cosmological constant problem,''
  Phys.\ Rev.\ D {\bf 65}, 125028 (2002)
  %%CITATION = doi:10.1103/PhysRevD.65.125028;%%
  %252 citations counted in INSPIRE as of 02 Jun 2017
%%%%%%%%%%%%%%%%%%%%%%%%%%%%%%%%%%%%%%%%%%%%%%%%%%%%%%%%%%%%
  \bibitem{flato}  F. Bayern, M. Flato, C. Fr\"onsdal, A. Lichnerowicz and D. Sternheimer, Ann.\,of Phys. {\bf 111}, 61 (1978), ibid.
  {\bf 111}, 151 (1978).
%%%%%%%%%%%%%%%%%%%%%%%%%%%%%%%%%%%%%%%%%%%%%%%%%%%%%%%%%%%%
%%%%%%%%%%%%%%%%%%%%%%%%%%%%%%%%%%%%%%%%%%%%%%%%%%%%%%%%%%%%
\bibitem{catta} A. S. Cattaneo, and D. Indelicato, Formality and Star Products, in {it Poisson Geometry, deformations quantization and Group Representation,} 2004,
Cambridge University Press.
%%%%%%%%%%%%%%%%%%%%%%%%%%%%%%%%%%%%%%%%%%%%%%%%%%%%%%%%%%%%
%%%%%%%%%%%%%%%%%%%%%%%%%%%%%%%%%%%%%%%%%%%%%%%%%%%%%%%%%%%%
\bibitem{wilde} M. De Wilde and P. B.  A. Lecomte,\,Lett.\,Math.\,Phys. {\bf 7},  487 (1994).
%%%%%%%%%%%%%%%%%%%%%%%%%%%%%%%%%%%%%%%%%%%%%%%%%%%%%%%%%%%%
\bibitem{fedosov} B.~v.~Fedosov,
  %``A Simple geometrical construction of deformation quantization,''
  J.\ Diff.\ Geom.\  {\bf 40}, no. 2, 213 (1994).
%%%%%%%%%%%%%%%%%%%%%%%%%%%%%%%%%%%%%%%%%%%%%%%%%%%%%%%%%%%%
%%%%%%%%%%%%%%%%%%%%%%%%%%%%%%%%%%%%%%%%%%%%%%%%%%%%%%%%%%%%
\bibitem{flato1} M. Flato, A. Lichnerowicz and D. Sternheimer, C.\,R.\,Sci.\,Paris, Ser. A-B, {\bf 283}, 61 (1976) and ibid, {\bf 283}, 111 (1976).
%%%%%%%%%%%%%%%%%%%%%%%%%%%%%%%%%%%%%%%%%%%%%%%%%%%%%%%%%%%%
%%%%%%%%%%%%%%%%%%%%%%%%%%%%%%%%%%%%%%%%%%%%%%%%%%%%%%%%%%%%
%%%%%%%%%%%%%%%%%%%%%%%%%%%%%%%%%%%%%%%%%%%%%%%%%%%%%%%%%%%%%
 \bibitem{kosen}  M.  Konsevitch, \,Lett. in Mathematical  Physics {\bf 66}, 157 (2003);
 see also, C. Esposito, \lq \lq Lectures on Deformation quantization of Poisson manifolds",  arXiv:1207.3287v2.
%%%%%%%%%%%%%%%%%%%%%%%%%%%%%%%%%%%%%%%%%%%%%%%%%%%%%%%%%%%%%
%%%%%%%%%%%%%%%%%%%%%%%%%%%%%%%%%%%%%%%%%%%%%%%%%%%%%%%%%%%

\bibitem{connes} A. Connes, Noncommutative Geometry, Pergamon (1994).
%%%%%%%%%%%%%%%%%%%%%%%%%%%%%%%%%%%%%%%%%%%%%%%%%%%%%%%%%%%%
%%%%%%%%%%%%%%%%%%%%%%%%%%%%%%%%%%%%%%%%%%%%%%%%%%%%%%%%%%%%
%%%%%%%%%%%%%%%%%%%%%%%%%%%%%%%%%%%%%%%%%%%%%%%%%%%%%%%%%%%%
%%%%%%%%%%%%%%%%%%%%%%%%%%%%%%%%%%%%%%%%%%%%%%%%%%%%%%%%%%%%
%%%%%%%%%%%%%%%%%%%%%%%%%%%%%%%%%%%%%%%%%%%%%%%%%%%%%%%%%%%%
%%%%%%%%%%%%%%%%%%%%%%%%%%%%%%%%%%%%%%%%%%%%%%%%%%%%%%%%%%%%
 \bibitem{marti}
  J.~Gamboa, C.~Ramirez and M.~Ruiz-Altaba,
  %``Null Spinning Strings,''
  Nucl.\ Phys.\ B {\bf 338}, 143 (1990); ibid, Phys.\ Lett.\ B {\bf 225} (1989) 335.
  %%%%%%%%%%%%%%%%%%%%%%%%%%%%%%%%%%%%%%%%%%%%%%%%%%%%%%%%%%%%%
%%%%%%%%%%%%%%%%%%%%%%%%%%%%%%%%%%%%%%%%%%%%%%%%%%%%%%%%%%%%%
 \bibitem{isham}C.~J.  Isham, Proc. \,R.  Soc. \, {\bf  351A} (1976)  209; J.~Gamboa and F.~Mendez,
%  %``Quantum gravity at very high-energies,'';
Nucl.\ Phys.\ B {\bf 600} (2001) 378; M.~Pilati,
  %``Strong Coupling Quantum Gravity. 1. Solution in a Particular Gauge,''
  Phys.\ Rev.\ D {\bf 26}, 2645 (1982); C.~J.~Isham,
  %``Canonical quantum gravity and the problem of time,''
  gr-qc/9210011; T.~Banks,
  %``T C P, Quantum Gravity, the Cosmological Constant and All That...,''
  Nucl.\ Phys.\ B {\bf 249} (1985) 332.
%
%%%%%%%%%%%%%%%%%%%%%%%%%%%%%%%%%%%%%%%%%%%%%%%%%%%%%%%%%%%%%
\bibitem{marc} M.~Henneaux,
  %``Geometry of Zero Signature Space-times,''
  Bull.\ Soc.\ Math.\ Belg.\  {\bf 31}, 47 (1979).
%%%%%%%%%%%%%%%%%%%%%%%%%%%%%%%%%%%%%%%%%%%%%%%%%%%%%%%%%%%%
   \bibitem{ncqm}   See   for   example,   J.~Gamboa,   M.~Loewe   and
     J.~C.~Rojas,
  %``Noncommutative quantum mechanics,''
  Phys.\   Rev.\   D   {\bf   64}  (2001)   067901;   V.~P.~Nair   and
  A.~P.~Polychronakos,
  %``Quantum mechanics on the noncommutative plane and sphere,''
  Phys.\  Lett.\ B  {\bf  505} (2001)  267; O.~Bertolami,  J.~G.~Rosa,
  C.~M.~L.~de Aragao, P.~Castorina and D.~Zappala,
  %``Noncommutative gravitational quantum well,''
  Phys.\ Rev.\ D {\bf 72}, 025010 (2005); J.~Gamboa, M.~Loewe, F.~Mendez and J.~C.~Rojas,
  %``The Landau problem and noncommutative quantum mechanics,''
  Mod.\ Phys.\ Lett.\ A {\bf 16}, 2075 (2001); J.~Gamboa, M.~Loewe, F.~Mendez and J.~C.~Rojas,
  %``The Landau problem and noncommutative quantum mechanics,''
  Mod.\ Phys.\ Lett.\ A {\bf 16}, 2075 (2001); H.~Falomir, J.~Gamboa, M.~Loewe, F.~Mendez and J.~C.~Rojas,
  %``Testing spatial noncommutativity via the Aharonov-Bohm effect,''
  Phys.\ Rev.\ D {\bf 66}, 045018 (2002); D.~Karabali, V.~P.~Nair and A.~P.~Polychronakos,
  %``Spectrum of Schrodinger field in a noncommutative magnetic monopole,''
  Nucl.\ Phys.\ B {\bf 627}, 565 (2002); C.~Acatrinei,
  %``Path integral formulation of noncommutative quantum mechanics,''
  JHEP {\bf 0109}, 007 (2001).
%%%%%%%%%%%%%%%%%%%%%%%%%%%%%%%%%%%%%%%%%%%%%%%%%%%%%%%%%%%%%
%%%%%%%%%%%%%%%%%%%%%%%%%%%%%%%%%%%%%%%%%%%%%%%%%%%%%%%%%%%%%
\bibitem{freidel1} L.~Freidel, R.~G.~Leigh and D.~Minic,
  %``Noncommutativity of closed string zero modes,''
  Phys.\ Rev.\ D {\bf 96}, no. 6, 066003 (2017).
%%%%%%%%%%%%%%%%%%%%%%%%%%%%%%%%%%%%%%%%%%%%%%%%%%%%%%%%%%%%%
\bibitem{freidel2}  L.~Freidel, R.~G.~Leigh and D.~Minic,
  %``Intrinsic non-commutativity of closed string theory,''
  JHEP {\bf 1709} (2017) 060.

%

    \bibitem{frieman} J.~Frieman, M.~Turner and D.~Huterer,
  %``Dark Energy and the Accelerating Universe,''
  Ann.\ Rev.\ Astron.\ Astrophys.\ {\bf 46} (2008) 385.

%%%%%%%%%%%%%%%%%%%%%%%%%%%%%%%%%%%%%%%%%%%%%%%%%%%%%%%%%%%%
%%%%%%%%%%%%%%%%%%%%%%%%%%%%%%%%%%%%%%%%%%%%%%%%%%%%%%%%%%%%
 \bibitem{perl}  S.~Perlmutter  {\it   et  al.}  [Supernova  Cosmology
   Project Collaboration],
  %``Measurements of Omega and Lambda from 42 high redshift supernovae,''
  Astrophys.\ J.\ {\bf 517}, 565 (1999).

%%%%%%%%%%%%%%%%%%%%%%%%%%%%%%%%%%%%%%%%%%%%%%%%%%%%%%%%%%%%
%%%%%%%%%%%%%%%%%%%%%%%%%%%%%%%%%%%%%%%%%%%%%%%%%%%%%%%%%%%%
  \bibitem{riess}  A.~G.~Riess {\it  et  al.}  [Supernova Search  Team
    Collaboration],
  %``Observational  evidence  from   supernovae  for  an  accelerating
  %universe and a cosmological constant,''
  Astron.\ J.\ {\bf 116} (1998) 1009.


%%%%%%%%%%%%%%%%%%%%%%%%%%%%%%%%%%%%%%%%%%%%%%%%%%%%%%%%%%%%
%%%%%%%%%%%%%%%%%%%%%%%%%%%%%%%%%%%%%%%%%%%%%%%%%%%%%%%%%%%%


%%%%%%%%%%%%%%%%%%%%%%%%%%%%%%%%%%%%%%%%%%%%%%%%%%%%%%%%%%%%
\bibitem{zak} J. Zak, \ Phys. \ Rev. {\bf 134}, A1602 (1964).
%%%%%%%%%%%%%%%%%%%%%%%%%%%%%%%%%%%%%%%%%%%%%
\bibitem{freidel3}   L.~Freidel, R.~G.~Leigh and D.~Minic,
  %``Quantum Spaces are Modular,''
  Phys.\ Rev.\ D {\bf 94}, 104052 (2016).

\bibitem{faddeev}  L.~D.~Faddeev,
  %``Operator Anomaly for the Gauss Law,''
  Phys.\ Lett.\  {\bf 145B}, 81 (1984).
  %%%%%%%%%%%%%%%
\bibitem{jackiw} R. Jackiw, \ Phys. \ Rev. Lett. {\bf 54}, 159 1985;  R. Jackiw, \ Phys. \ Lett. {\bf B154}, 303 (1985); B. Grossman, \ Phys. \ Lett. {B152}, 93 (1985).
\end{thebibliography}
\end{document}